\documentclass[10pt,journal,compsoc]{IEEEtran}

\usepackage{amsmath,amssymb,amsfonts}
\usepackage{algpseudocode}
\usepackage{algorithmicx,algorithm}
\usepackage{algpseudocode}
\usepackage{amsmath}
\usepackage{stfloats}
\usepackage{graphicx}
\usepackage{booktabs}
\usepackage{textcomp}
\usepackage{xcolor}
\usepackage{bm}
\usepackage{subfigure}
\usepackage{authblk}
\usepackage{array}
\usepackage{diagbox}
\usepackage{multirow}
\usepackage{url}
\usepackage{indentfirst}
\usepackage{breqn}
\usepackage{tabularx}
\usepackage{makecell}
\usepackage{ragged2e}

\usepackage{tabularx}
\newcolumntype{Y}{>{\centering\arraybackslash}X}

\usepackage{caption}
\newcolumntype{C}[1]{>{\centering}p{#1}}
\begin{document}


\title{Autonomous Collaborative Scheduling of Time-dependent UAVs, Workers and Vehicles for Crowdsensing in Disaster Response\\}

\author{Lei~Han*, Yitong~Guo*, Pengfei~Yang, Zhiyong~Yu, Liang~Wang, Quan~Wang and~Zhiwen~Yu,~\IEEEmembership{Senior~Member~IEEE}
	\thanks{Lei Han, Yitong Guo, Pengfei Yang (corresponding author) and Quan Wang are with School of Computer Science and Technology, Xidian University, Xi'an, China. E-mail: \{hanlei@xidian.edu.cn; yt\_guo@stu.xidian.edu.cn; pfyang@xidian.edu.cn; qwang@xidian.edu.cn\}}
	\thanks{Zhiyong Yu is with College of Computer and Data Science, Fuzhou University, Fuzhou, China. E-mail: yuzhiyong@fzu.edu.cn.}
	\thanks{Liang Wang is School of Computer Science, Northwestern Polytechnical University, Xi'an, China. E-mail: liangwang@nwpu.edu.cn.}
	\thanks{Zhiwen Yu is with Harbin Engineering University, Harbin, China. E-mail: zhiwenyu@nwpu.edu.cn.}
	\thanks{*These authors contributed equally to this work.}}

\markboth{}%
{}

\IEEEtitleabstractindextext{%
\begin{abstract}
\justifying 
Natural disasters have caused significant losses to human society, and the timely and efficient acquisition of post-disaster environmental information is crucial for the effective implementation of rescue operations. Due to the complexity of post-disaster environments, existing sensing technologies face challenges such as weak environmental adaptability, insufficient specialized sensing capabilities, and limited practicality of sensing solutions. This paper explores the heterogeneous multi-agent online autonomous collaborative scheduling algorithm HoAs-PALN, aimed at achieving efficient collection of post-disaster environmental information. HoAs-PALN is realized through adaptive dimensionality reduction in the matching process and local Nash equilibrium game, facilitating autonomous collaboration among time-dependent UAVs, workers and vehicles to enhance sensing scheduling. (1) In terms of adaptive dimensionality reduction during the matching process, HoAs-PALN significantly reduces scheduling decision time by transforming a five-dimensional matching process into two categories of three-dimensional matching processes; (2) Regarding the local Nash equilibrium game, HoAs-PALN combines the softmax function to optimize behavior selection probabilities and introduces a local Nash equilibrium determination mechanism to ensure scheduling decision performance. Finally, we conducted detailed experiments based on extensive real-world and simulated data. Compared with the baselines (GREEDY, K‑WTA, MADL and MARL), HoAs‑PALN improves task completion rates by 64.12\%, 46.48\%, 16.55\%, and 14.03\% on average, respectively, while each online scheduling decision takes less than 10 seconds, demonstrating its effectiveness in dynamic post‑disaster environments.
\end{abstract}

}

\maketitle

\section{Introduction}
\label{section 1}
According to the 2023 Global Nature Disaster Assessment Report \cite{1-gddat} and statistics from the United Nations Office for Disaster Risk Reduction \cite{2-preventionweb}, frequent large-scale natural disasters have resulted in over 93.05 million people affected globally each year, with direct economic losses reaching as high as \$202.652 billion. To mitigate the severe impacts of natural disasters, it is crucial to organize post-disaster relief efforts quickly and effectively. Taking earthquake disasters as an example \cite{3-Earthquake}, the survival probability of survivors is approximately 90\% on the first day following the earthquake, but this significantly decreases to around 50\%-60\% by the second day. In such emergency situations, it is essential to obtain real-time and accurate information about the affected areas—including the extent of damage, severity, and potential hazards. This information supports the scientific formulation of rescue plans, the rational allocation of relief supplies, and the effective avoidance of safety risks.

Currently, there are two mainstream methods for collecting environmental information: sensor network sensing \cite{4-kandris2020applications}\cite{5-majid2022applications} and mobile crowdsensing \cite{6-ma2023qun}\cite{7-yu2021crowdsensing}. However, existing technologies in both methods struggle to collect post-disaster environmental information in real-time and with precision.Sensor network sensing relies on various pre-deployed fixed devices to gather environmental data. However, its low adaptability to complex, dynamic environments and its high dependence on infrastructure such as electricity and networks limit its effectiveness in post-disaster scenarios. On the other hand, mobile crowdsensing primarily uses portable smart devices as basic sensing units, relying on a central server for the scheduling of these units, which poses challenges in post-disaster contexts. The limitations are twofold: first, the complex conditions of post-disaster environments restrict human mobility, and the sensing capabilities of portable smart devices are often inadequate; second, damage to infrastructure such as electricity and networks can hinder the scheduling of sensing units that depend on a central server.

In recent years, with the proliferation and application of UAVs, their role in assisting post-disaster relief has become increasingly significant. UAVs offer advantages such as rapid deployment, high mobility, and the ability to flexibly carry specialized sensing equipment, prompting many researchers to explore their application in post-disaster environmental data collection \cite{8-wang2021energy}\cite{9-dai2022aoi}. However, research on the use of UAVs for post-disaster environmental data collection is based on two overly idealized assumptions, which hinder their practical application in real disaster environments. First, it is assumed that UAVs can autonomously execute data collection tasks. The low-altitude environment after a disaster is highly complex, and many data collection tasks require reasonable obstacle avoidance and precise targeting. Without professional operators, relying solely on UAVs for autonomous control to complete data collection tasks is impractical. Second, it is assumed that UAVs can automatically recharge at charging stations. Currently, there are very few charging stations in urban areas that provide services for UAVs, and the limited number of charging stations available post-disaster may be severely damaged and unusable. Additionally, in outdoor environments, UAVs face significant challenges in autonomously connecting to power sources without assistance.

To address these two major issues, recent research \cite{10-han2024collaborative} has attempted to leverage the differences and complementarities among UAVs, operators, and supply vehicles to efficiently complete post-disaster environmental data collection tasks. Specifically, workers can use portable terminals to control UAVs for agile data collection; supply vehicles can serve as mobile power sources to provide reliable charging services for low-battery UAVs. However, this work still has three shortcomings: first, due to the extreme randomness of natural disasters, obtaining reliable historical data post-disaster is extremely difficult, making solutions based on centralized training models challenging to apply in real post-disaster scenarios. Second, the proposed solutions are only applicable to fixed coupling relationships among UAVs, operators, and supply vehicles, whereas their coupling relationships in post-disaster environments are often time-dependent. Third, the proposed solutions require broadcasting all attributes of agents and global environmental features when scheduling UAVs, operators, and supply vehicles, while in reality, different agents can often only access information within their limited communication range.

\begin{figure}[ht]
	\setlength{\abovecaptionskip}{0.2cm}
	\setlength{\belowcaptionskip}{-0.25cm}
	\centering 
	\includegraphics[width=\linewidth]{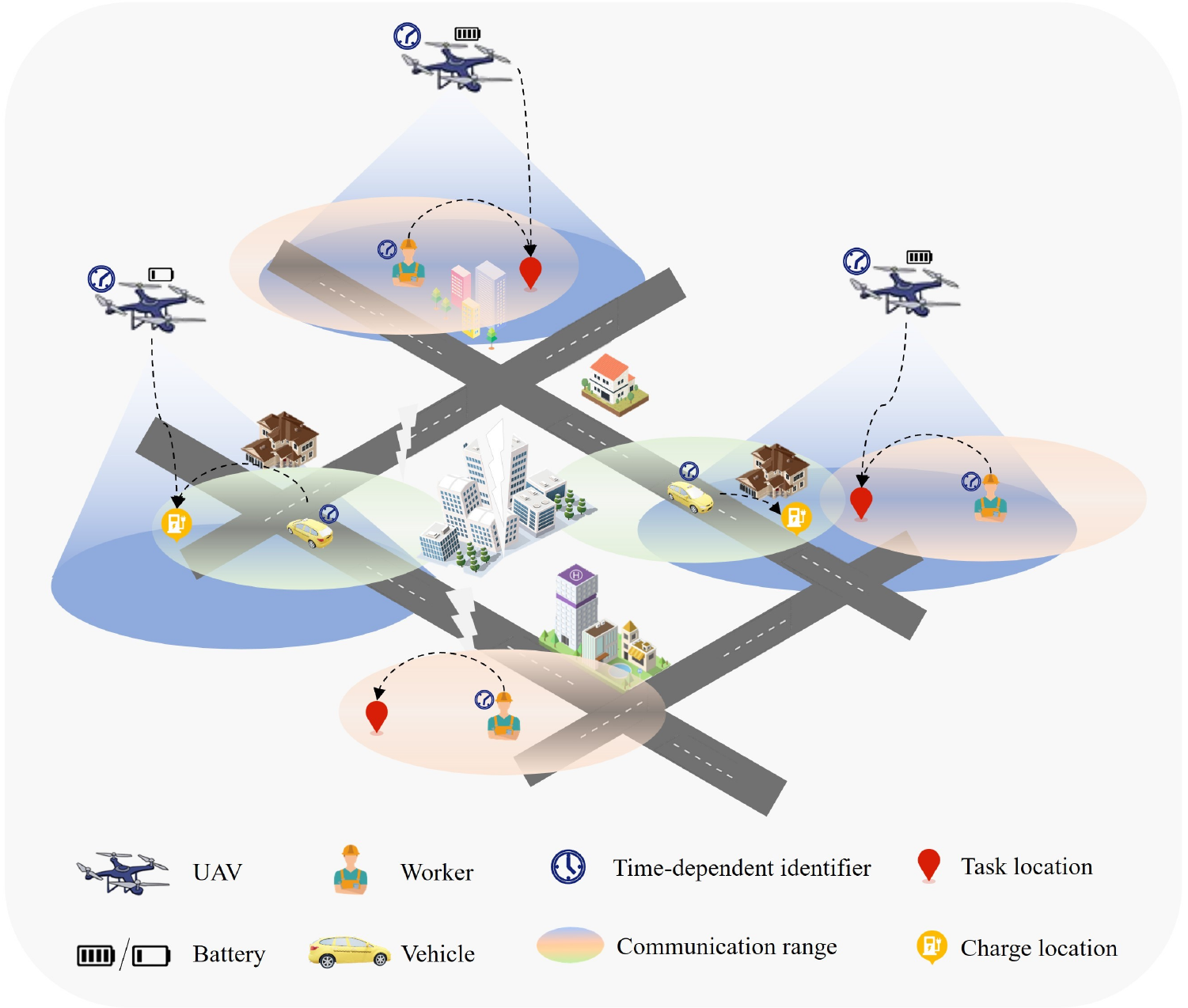}
	\caption{An example of time-dependent autonomous collaboration among UAVs, workers and vehicles.}
	\label{figure1}
\end{figure}

Therefore, we investigate the autonomous collaborative scheduling of time-dependent UAVs, workers and vehicles for crowdsensing in post-disaster responses, as illustrated in Fig. \ref{figure1}. In our research, three objectives need to be achieved. First, we aim to propose an online heterogeneous multi-agent collaborative scheduling algorithm that does not require training based on historical data. Second, the proposed algorithm should effectively schedule time-dependent UAVs, workers, and vehicles to efficiently complete post-disaster data collection tasks. Third, all UAVs, workers, and vehicles should make autonomous scheduling decisions based solely on information within their own communication range, striving for optimal outcomes. To achieve these goals, we face three major challenges: (1) The complexity of the five-dimensional matching process (UAVs, workers, vehicles, task points, and charge points) and the excessive computational resource consumption make it difficult to implement in resource-constrained autonomous scheduling scenarios, necessitating the design of dimensionality reduction methods to simplify the matching process; (2) Our research involves resource competition games among homogeneous agents and cooperative games among heterogeneous agents, requiring agents to consider constraints related to their own communication, mobility, and functionality while addressing the complex interactions arising from the games; (3) In communication-constrained scenarios, heterogeneous multi-agents must make autonomous collaborative scheduling decisions while ensuring that the scheduling system achieves overall relative stability, which necessitates effective determination of whether the collaborative scheduling results reach Nash equilibrium.

To address the aforementioned challenges, we propose a heterogeneous multi-agent online autonomous collaborative scheduling algorithm based on probabilistic action selection and local Nash game, termed HoAs-PALN. First, to tackle the high complexity of five-dimensional matching and the substantial computational load of the matching process, we introduce a dimensionality reduction method based on the expected benefits of UAVs (performing tasks or being charged). This method divides the five-dimensional matching problem of ``UAV-worker-vehicle-task point-charge point'' into two three-dimensional matching categories: ``UAV-worker-task point'' and ``UAV-vehicle-charge point'', significantly reducing the computational complexity of the matching process. Second, to address the complex coexistence of cooperation and competition among multiple agents, as well as the stability determination of the scheduling system in communication-constrained scenarios, we design a heterogeneous multi-agent autonomous collaborative scheduling algorithm. This algorithm employs the softmax function to convert distances into action selection probabilities, assigning higher selection probabilities to superior actions to enhance agent exploration efficiency and improve algorithm convergence speed. Additionally, we propose a method for determining the local Nash equilibrium state of agents, ensuring that the scheduling results allow each agent to achieve a local Nash equilibrium within its communication coverage area.

In summary, this work makes the following contributions:

(1) To the best of our knowledge, this is the first study to investigate the autonomous collaborative scheduling of time-dependent UAVs, workers, and vehicles for crowdsensing in disaster response. In addition, we prove that the problem is NP-Hard.

(2) To tackle the aforementioned issues, we propose the HoAs-PALN algorithm for heterogeneous multi-agent online autonomous collaborative scheduling. HoAs-PALN is realized through probabilistic action selection and local Nash games, comprising two main components: 1) dimensionality reduction of the matching process based on the expected benefits of UAVs; and 2) autonomous collaborative scheduling of heterogeneous multi-agents based on probabilistic action selection and local Nash equilibrium determination.

(3) We conducted detailed experiments based on extensive real-world and simulated data. Compared with baseline algorithms including GREEDY, K-WTA, MADL, MARL, and HoAs-RALN, the proposed HoAs-PALN achieves significant improvements in task completion rates. Moreover, across all test datasets, the time taken for a single matching decision by HoAs-PALN does not exceed 10 seconds. To facilitate reproducibility, all the source code and experimental datasets have been made publicly available at: [GitHub Link] \cite{github}.

The contents of this paper are arranged as follows: Section \ref{section 2} discusses some related works; Section \ref{section 3} formulates the problem of autonomous collaborative scheduling of time-dependent UAVs, workers, and vehicles for crowdsensing in disaster responses; Section \ref{section 4} provides a comprehensive and detailed explanation of the HoAs-PALN method; Section \ref{section 5} presents detailed experiments; and finally, Section \ref{section 6} summarizes the conclusions and outlines future work.

\section{Related Work}
\label{section 2}
The attributes and functions of heterogeneous agents such as UAVs, vehicles, and humans differ significantly, leading to substantial variations in heterogeneous multi-agent collaborative modes. Existing research on heterogeneous multi-agent collaboration can be categorized into three types: (1) collaboration based on spatial position division; (2) collaboration based on task phase division; and (3) collaboration based on fine-grained coordination.

\subsection{Collaboration Based on Spatial Position Division}
\label{section 2.1}
This collaborative mode can be further divided into three subcategories. The most common mode involves the collaboration between UAVs and vehicles in terms of spatial positioning. For instance, Wu et al. \cite{11-wu2020cooperative} planned joint paths for UAVs and vehicles to minimize overall monitoring time through differentiated deployment of aerial and ground sensing units. In the field of navigation, Zuo et al. \cite{12-zuo2023real} utilized UAVs to collect ground information from the air, thereby enhancing vehicle navigation capabilities in complex terrestrial environments. Zang et al. \cite{13-zang2024coordinated} aimed to use aerial data collected by UAVs to reduce collision risks among large ground vehicle clusters. Research has also explored the collaborative construction of sensing maps by employing UAVs to provide wide-area aerial imagery while vehicles supplement ground details \cite{14-christie2017radiation}\cite{15-michael2014collaborative}\cite{16-kalaitzakis2021marsupial}. Additionally, some studies have focused on the spatial collaboration between UAVs/vehicles and humans. For example, Ding et al. \cite{17-ding2021crowdsourcing} developed a low-cost urban sensing system based on the spatial distribution differences between human groups and autonomous vehicles. Zheng et al. \cite{18-zheng2019evolutionary} investigated how to utilize human groups and UAVs to simultaneously search different spatial areas for efficiently locating fugitives. In emergency response scenarios, Minaeian et al. \cite{19-minaeian2015vision} examined how high-altitude UAVs can assist ground human patrols by providing real-time feedback on suspicious target locations. Finally, a few studies have explored collaborative spatial positioning among UAVs, vehicles, and humans. For instance, Wu et al. \cite{20-wu2022task} coordinated small unmanned rotorcraft, intelligent connected vehicles, and human groups to different areas to execute multiple ground target searches.

\subsection{Collaboration Based on Task Phase Division}
\label{section 2.2}
Similarly, this collaborative mode also encompasses three main types. The most prevalent mode involves collaboration between UAVs and vehicles based on task phases. For example, studies \cite{21-russell2016artificial}\cite{22-plocher2017german} utilized UAVs for rapid area exploration and point-of-interest marking, followed by employing autonomous vehicles for detailed target detection. To address the issue of limited UAV endurance, Wang et al. \cite{23-wang2022mobile} researched how vehicles equipped with wireless charging devices could provide on-demand, safe, and efficient charging services to UAVs. Furthermore, Wang et al. \cite{24-wang2023air} considered how vehicles could serve as carriers for UAVs to extend their range for long-duration monitoring tasks. Leveraging the strong transportation efficiency of trucks alongside the precise delivery capabilities of UAVs, Luo et al. \cite{25-luo2024collaborative} developed a two-phase logistics delivery system involving trucks and UAVs. Similarly, Fu et al. \cite{26-fu2022energy} utilized UAVs for data collection while ground vehicles transmitted the collected data, achieving a balance between efficient data collection and processing. Some research has also focused on collaboration between UAVs/vehicles and humans based on task phases. For instance, Niroui et al. \cite{27-niroui2019deep} employed UAVs to capture aerial images of disaster areas, which were then transmitted in real-time to human operators for key target annotation via augmented reality (AR). In construction engineering, research \cite{28-krizmancic2020cooperative} utilized UAVs for high-altitude scanning tasks, followed by human operators performing semantic analysis and material transportation. In power maintenance scenarios, Zheng et al. \cite{29-zheng2020evolutionary} employed UAVs to inspect power transmission faults and relay the fault information to operators for facility repair. To facilitate the rapid search for missing tourists, Xu et al. \cite{30-xu2024iterated} established a two-phase model involving UAV detection and search-and-rescue operations, significantly improving rescue efficiency. Lastly, a few studies have explored collaboration among UAVs, vehicles, and humans based on task phases. For example, Ghassemi et al. \cite{31-ghassemi2019decentralized} attempted to construct a three-phase collaborative model (e.g. UAVs rapidly scanning disaster areas, vehicles transporting supplies, and humans conducting centralized command and scheduling) to efficiently complete disaster response tasks.

\subsection{Collaboration Based on Fine-grained Coordination}
\label{section 2.3}
This collaborative mode also generally involves three types. The most common collaborative mode is fine-grained coordination between UAVs and vehicles. Hu et al. \cite{32-hu2023collaboration} proposed a 3D collaborative detection framework (CoCa3D), which enables real-time sharing of critical data between UAVs and vehicles through deep uncertainty sensing. Wang et al. \cite{33-wang2022multi} jointly optimized the motion trajectories of UAVs and vehicles to maintain target tracking lines, addressing the issue of obstacles obstructing UAV-based target monitoring in urban environments. Additionally, some research has focused on fine-grained coordination between UAVs/vehicles and humans. To tackle the challenges of communication redundancy and transmission delays that complicate collaborative sensing among heterogeneous agents, research \cite{34-yang2023how2comm} utilized semantic communication via UAVs to transmit high-quality information features to humans, supporting real-time collaborative decision-making. Moreover, research \cite{35-manjunatha2020using} innovatively integrated brain-computer interfaces into air-ground swarm control processes, achieving dynamic mapping of human intentions to UAV behaviors. Finally, a few studies have explored fine-grained coordination among UAVs, vehicles, and humans. In recent research, Miller et al. \cite{36-miller2022stronger} designed an LLM-driven air-ground swarm coordination system that dynamically adjusts task priorities through natural language interactions among UAVs, unmanned vehicles, and humans. In terms of data collection, Han et al. \cite{10-han2024collaborative} explored collaborative path planning methods for UAVs, vehicles, and human groups to achieve precise coordination in post-disaster response scenarios.

\subsection{Summary}
\label{section 2.4}
The primary issues with the aforementioned research include the lack of synergy in collaboration modes based on spatial position division and task phase division. These modes often decompose large tasks into smaller, phased tasks assigned to different agents or allocate tasks to different regions based on agents' mobility capabilities. In contrast, fine-grained coordination among heterogeneous multi-agents can fully leverage their functional characteristics, maximizing complementary advantages. However, the reasoning capabilities of large language models often struggle with highly complex problems. Methods based on deep learning and reinforcement learning require extensive historical data for model training, but obtaining such data in disaster scenarios, which are characterized by extreme randomness, is challenging. Furthermore, deep learning and reinforcement learning methods exhibit strong dependencies on the coupling relationships among multi-agents \cite{37-li2024reinforcement}, leading to significant declines in decision-making performance if the number of agents or their online states change. Therefore, further in-depth research is urgently needed on the autonomous collaborative scheduling of time-dependent UAVs, workers and vehicles for crowdsensing in disaster responses.

\section{Problem Formulation}
\label{section 3}

In this section, we first clarify the definitions of several fundamental concepts, followed by an analysis of the constraints and optimization objectives related to the problem under investigation.

\textbf{{\itshape Definition 1.}} The set of UAVs is denoted as $UAV=\{ ua{v_0},...ua{v_i},...\}$, where $uav_i = \langle uLoc_i\allowbreak, uRge_i\allowbreak, Fullpower_i\allowbreak, uPow_i\allowbreak, U\_Radius_i\allowbreak, U\_uptime_i\allowbreak, U\_downtime_i \rangle$ represents the i-th UAV in the set. $uLo{c_i}$ denotes the current location of $ua{v_i}$; $uRg{e_i}$ indicates the movement speed of $ua{v_i}$; $Fullpowe{r_i}$ represents the maximum distance that $ua{v_i}$ can travel when fully charged; $uPo{w_i} \in \left[ {0,Fullpowe{r_i}} \right]$ specifies the distance that $ua{v_i}$ can travel based on its current battery level; $U\_Radiu{s_i}$ indicates the communication range of $ua{v_i}$; $U\_uptim{e_i}$ and $U\_downtim{e_i}$ represent the uptime and downtime of $ua{v_i}$, respectively, which are used to simulate the time-dependent behavior of $ua{v_i}$.

\textbf{{\itshape Definition 2.}} The set of workers is denoted as $Worker=\{ worker _0,...worker _j,...\} $, where $worker_j=\langle wLoc_j\allowbreak, wRge_j\allowbreak, W\_Radius_j\allowbreak, W\_uptime_j\allowbreak,W\_downtime_j \rangle$ represents the j-th worker in the set. $wLo{c_j}$ denotes the current location of $wor{ker _j}$; $wRg{e_j}$ indicates the movement speed of $wor{ker _j}$; $W\_Radiu{s_j}$ indicates the communication range of $wor{ker _j}$; $W\_uptim{e_j}$ and $W\_downtim{e_j}$ represent the uptime and downtime of $wor{ker _j}$, respectively, which are used to simulate the time-dependent behavior of $wor{ker _j}$.

\textbf{{\itshape Definition 3.}} The set of vehicles is denoted as $Vehicle=\{ vehicle_0,...,vehicle_k,...\}$, where $vehicle_k=\langle vLoc_k\allowbreak, vRge_k\allowbreak, V\_Radius_k\allowbreak, V\_chargePow_k\allowbreak, V\_uptime_k\allowbreak, V\_downtime_k\rangle $ represents the k-th vehicle in the set. $vLo{{c}_{k}}$ denotes the current location of $vehicl{{e}_{k}}$; $vRg{{e}_{k}}$ indicates the movement speed of $vehicl{{e}_{k}}$; $V\_Radiu{{s}_{k}}$ indicates the communication range of $vehicl{{e}_{k}}$; $V\_chargePo{{w}_{k}}$ represents the charging power provided by $vehicl{{e}_{k}}$ to a UAV per unit time, which corresponds to the UAV's endurance enhancement; $V\_uptim{{e}_{k}}$ and $V\_downtim{{e}_{k}}$ represent the uptime and downtime of $vehicl{{e}_{k}}$, respectively, which are used to simulate the time-dependent behavior of $vehicl{{e}_{k}}$.

\textbf{{\itshape Definition 4.}} The set of task points is denoted as $Task=\left\{ tas{{k}_{0}},...,tas{{k}_{x}},... \right\}$, where $task_x=\langle tLoc_x\allowbreak ,T\_costPow_x \rangle $ represents the x-th task point in the set. $tLo{{c}_{x}}$ denotes the location of $tas{{k}_{x}}$; and $T\_costPo{{w}_{x}}$ indicates the distance a UAV needs to travel in order to complete $tas{{k}_{x}}$.

\textbf{{\itshape Definition 5.}} The set of charge points is denoted as $Charge=\langle charg{{e}_{0}},...,charg{{e}_{y}},... \rangle $, where $charg{{e}_{y}}=\langle chLo{{c}_{y}} \rangle $ represents the y-th charge point in the set. $chLo{{c}_{y}}$ denotes the location of $charg{{e}_{y}}$.

Before defining the problem in this paper, we need to introduce the following constraints:

(1) When $ua{v_i}$ and $wor{ker _j}$ meet at location $tLo{{c}_{x}}$, $tas{{k}_{x}}$ can be completed, which causes the $uPo{w_i}$ of $ua{v_i}$ to decrease by $T\_costPo{{w}_{x}}$, see Equation (\ref{equation1}).
\begin{gather}
	task_x = \emptyset \; \text{and} \; uPow_i = uPow_i - T\_costPo{{w}_{x}},\notag
	\\ \text{if} \; uLoc_i = wLoc_j = tLoc_x
	\label{equation1}
\end{gather}

(2) When $ua{v_i}$ and $wor{ker _j}$ are executing $tas{{k}_{x}}$, the time required for both $ua{v_i}$ and $wor{ker _j}$ to complete the task is $T\_costPo{{w}_{x}}/uRg{e_i}$.

(3) When $ua{v_i}$ and $vehicl{{e}_{k}}$ meet at $charg{{e}_{y}}$, $vehicl{{e}_{k}}$ can charge $ua{v_i}$ until it is fully charged, see Equation (\ref{equation2}).
\begin{gather}
	uPow_i = Fullpower_i,\; \text{if} \; uLoc_i=vLoc_k=chLoc_y
	\label{equation2}
\end{gather}

(4) When $vehicl{{e}_{k}}$ charges $ua{v_i}$ at $charg{{e}_{y}}$, the time required for both $vehicl{{e}_{k}}$ and $ua{v_i}$ to complete the charging process is $(Fullpowe{r_i}-uPo{w_i})/V\_chargePo{w_k}$. Note: If there are multiple UAVs waiting to be charged, $vehicl{{e}_{k}}$ will charge them in a first-come-first-served manner.

(5) $ua{v_i}$,  $wor{ker _j}$, and $vehicl{{e}_{k}}$ can obtain the current state information of all UAVs, workers, vehicles, task points, and charge points within their communication range.

(6) When $ua{v_i}$ chooses $charg{{e}_{y}}$, it must be ensured that the $uPo{w_i}$ of $ua{v_i}$ is sufficient to reach $chLo{{c}_{y}}$, see Equation (\ref{equation3}).
\begin{gather}
	Dis(chLoc_y,uLoc_i) \le uPow_i
	\label{equation3}
\end{gather}
Note: \text{Dis()} denotes the function to calculate the distance between two points.

(7) When $ua{v_i}$ chooses to execute $tas{{k}_{x}}$, it must be ensured that the $uPo{w_i}$ of $ua{v_i}$ is sufficient to reach $tLo{{c}_{x}}$, and after completing the task, $ua{v_i}$ must have enough power to return to the nearest charge point $Opt\_charg{{e}_{y}} \in Charge$, see Equation (\ref{equation4}).
\begin{gather}
	Dis(uLoc_i,tLoc_x)+T\_costPow_x+ \notag \\ 
	Dis(tLoc_x,Opt\_charge_y)  \le uPow_i
	\label{equation4}
\end{gather}

\textbf{Problem 0} (Autonomous Collaborative Scheduling of Time-dependent UAVs, Workers and Vehicles for Crowdsensing): Given $UAV$, $Worker$, $Vehicle$, $Task$ and $Charge$, under the constraints mentioned above, the $\{ua{v_i\}}$, $\{wor{ker _j}\}$, and $\{vehicl{{e}_{k}}\}$ will execute their actions $\{ A\_uav_i^{t \times Interval}\}$, $\{ A\_worker_j^{t \times Interval}\}$, and $\{ A\_vehicle_k^{t \times Interval}\}$ at moments $t \times Interval,(t=0,1,2,...)$. The objective is to maximize the number of tasks completed before $LimitTime$.
\begin{gather*}
	\textbf{confirm} \; \{...\{A\_uav_i^{t \times Interval}\}...\},	\{...\{ A\_worker_j^{t \times Interval}\}...\},\\\{...\{ A\_vehicle_k^{t \times Interval}\}...\}, \\
	(t=0,1,2,...) \; \text{and} \; t \times Interval \in [0, LimitTime] \\
	\textbf{max} \;\; Cplt\_Tasks \\ 
	s.t. \; \text{constraints} \; (1)(2)(3)(4)(5)(6)(7)
\end{gather*}

Each agent needs to make a decision at the same time. Therefore, it is not easy to establish a direct relationship between the total number of tasks completed within $LimitTime$ and the single decision of each agent. We can convert the optimization objective from ``$Cplt\_Tasks$'' to perceived task completion amount of single self-scheduling decision `` $Cplt\_Tasks\_Per^{t \times Interval},(t=0,1,2,...)$'', so we get \textbf{Problem 1}: 
\begin{gather*}
	\textbf{confirm} \; \{A\_uav_i^{t \times Interval}\},	\{ A\_worker_j^{t \times Interval}\},\\ \{ A\_vehicle_k^{t \times Interval}\}, \\
	(t=0,1,2,...) \; \text{and} \; t \times Interval \in [0, LimitTime] \\
	\textbf{max} \;\; Cplt\_Tasks\_Per^{t \times Interval} \\
	s.t.\; \text{constraints} \; (1)(2)(3)(4)(5)(6)(7)
\end{gather*}

\textbf{Lemma 1.} \textbf{Problem 1} is NP-Hard.

$Proof$: 
First, we make the following assumptions to simplify \textbf{Problem 1}:

(1) UAVs have unlimited power without considering constraints related to vehicles and charge points;

(2) UAVs are capable of autonomously completing tasks without considering constraints related to workers;

(3) All UAVs have an infinite communication ra\textbf{}nge, allowing for global communication coverage;

(4) When a UAV arrives at the location $tLo{{c}_{x}}$ of $tas{{k}_{x}}$, $tas{{k}_{x}}$ can be instantly completed without any additional expenditure, see Equation (\ref{equation5}).
\begin{gather}
	task_x = \emptyset ,\; \text{if} \; uLoc_i = wLoc_j = tLoc_x
	\label{equation5}
\end{gather}

Based on these assumptions, \textbf{Problem 1} can be simplified to \textbf{Problem 2}:

\begin{gather*}
	\textbf{confirm} \; \{A\_uav_i^{t \times Interval}\}, \\
	(t=0,1,2,...) \; \text{and} \; t \times Interval \in [0, LimitTime] \\
	\textbf{max} \;\; Cplt\_Tasks\_Per^{t \times Interval} \\
	s.t. \; \text{assumption} \; (4)
\end{gather*}

\textbf{Problem 2} is a special case of \textbf{Problem 1} with simpler constraints. By proving that \textbf{Problem 2} is NP-Hard, we can then utilize problem reduction to conclude that \textbf{Problem 1} is also NP-Hard. \textbf{Problem 2} explores how multiple UAVs can make a single self-scheduling decision to determine a one-to-one match with multiple tasks, aiming to maximize the volume of tasks completed in a single decision interval $Cplt\_Tasks\_Per^{t \times Interval},(t=0,1,2,...)$. From the description above, it is evident that \textbf{Problem 2} is a typical optimal subset selection problem. Given that the optimal subset selection problem is NP-Hard \cite{38-nemhauser1978analysis}, \textbf{Problem 1} is also NP-Hard.

\section{Methodology}
\label{section 4}

\begin{figure}[b]
	\setlength{\abovecaptionskip}{0.2cm}
	\setlength{\belowcaptionskip}{-0.25cm}
	\centering 
	\includegraphics[width=8.5cm]{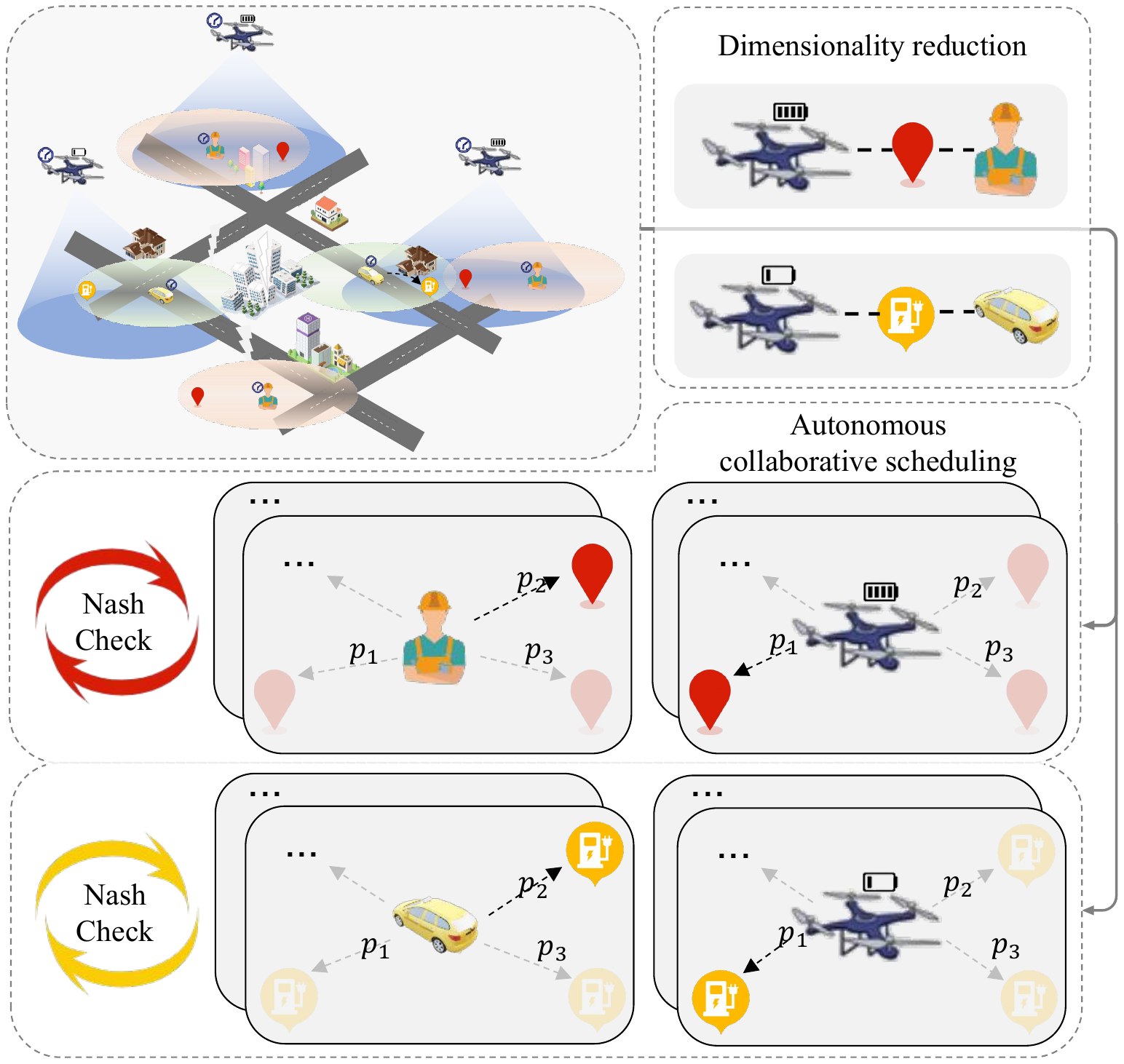}
	\caption{Framework of HoAs-PALN.}
	\label{figure2}
\end{figure}

To address \textbf{Problem 1}, we propose a Heterogeneous Multi-Agent Online Autonomous Collaborative Scheduling Algorithm Based on Probabilistic Action Selection and Local Nash Game, denoted as HoAs-PALN, whose framework is illustrated in Fig. \ref{figure2}. HoAs-PALN is generally divided into two parts. (1) Dimensionality Reduction of the Matching Process Based on the Estimated Expected Benefits of UAVs: The five-dimensional matching problem involving “UAVs–\allowbreak workers–\allowbreak vehicles–\allowbreak task points–\allowbreak charge points” is decomposed into two three-dimensional matching problems: “UAVs–\allowbreak workers–\allowbreak task points” and “UAVs–\allowbreak vehicles–\allowbreak charge points.” (2) Autonomous Collaborative Scheduling for Heterogeneous Multi-Agents Based on Probabilistic Action Selection and Local Nash Equilibrium Determination: A softmax function is employed to transform distances into action selection probabilities, assigning higher selection probabilities to superior actions. This approach increases the agents’ exploration efficiency and accelerates the convergence of the algorithm. In addition, a method is proposed to determine whether each agent has reached a local Nash equilibrium within its communication range, thereby ensuring that the system’s scheduling results achieve Nash equilibrium for every agent.

\subsection{Dimensionality Reduction of the Matching Process Based on the Expected Benefits Estimation for UAVs}
\label{section 4.1}

\begin{figure}[ht]
	\setlength{\abovecaptionskip}{0.2cm}
	\setlength{\belowcaptionskip}{-0.25cm}
	\centering 
	\includegraphics[width=\linewidth]{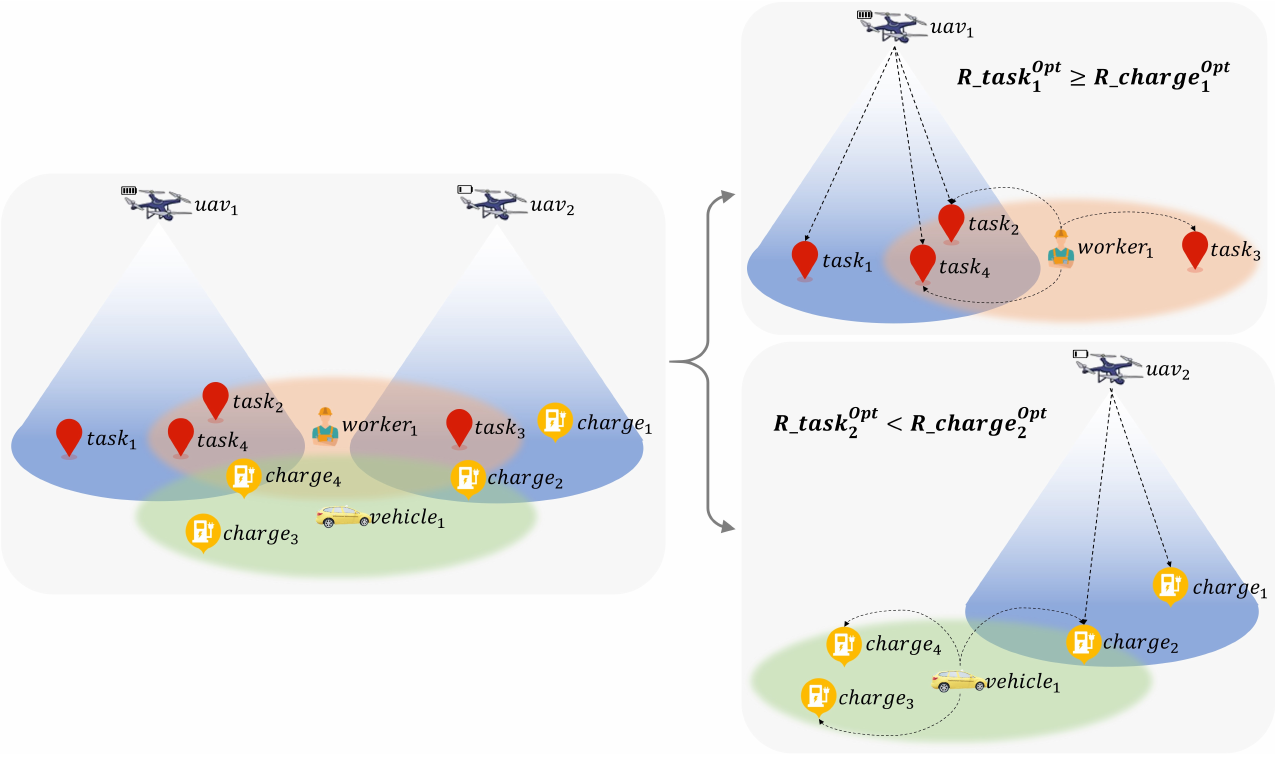}
	\caption{Schematic diagram of dimensionality reduction in the matching process.}
	\label{figure3}
\end{figure}

In order to reduce the computational complexity of the matching process, it is necessary to reduce the five-dimensional matching problem (UAVs–workers–vehicles–task points–charge points) to a three-dimensional matching problem (“UAVs–workers–task points” or “UAVs–vehicles–charge points”), as illustrated in Fig. \ref{figure3}. Notably, workers do not need to establish matching relationships with vehicles and charge points, and vehicles do not need to establish matching relationships with workers and task points. In other words, the autonomous scheduling processes for workers and vehicles are inherently three-dimensional and do not require dimensionality reduction in the matching process. The autonomous dimensionality reduction for the UAV matching process comprises three parts: (1) An estimation of the expected benefit when UAVs execute tasks; (2) An estimation of the expected benefit when UAVs are charged; (3) Proof of the reasonableness of the expected benefit comparison of UAVs. Note: The current system time is denoted by SysTime.

\subsubsection*{(1) An estimation of the expected benefit when UAVs execute tasks:}
	
Step 1. $ua{v_i}$ calculates the distance between it and the closest task point ($task_x^{Opt} \in Task$) within its communication range, denoted as $Dis(uLoc_i,tLoc_x^{Opt})$.

Step 2. $ua{v_i}$ calculates the reduction in its remaining energy ($\Delta uPow{T}_i$) and the required flight time ($\Delta uTime{T}_i$) for moving from $uLoc_i$ to ${tLoc}_x^{Opt}$, see Equation (\ref{equation6})(\ref{equation7}).
\begin{gather}
	\Delta uPowT_i=Dis(uLoc_i,tLoc_x^{Opt})
	\label{equation6}
\end{gather}
\begin{gather}
	\Delta uTimeT_i = \frac{Dis(uLoc_i,tLoc_x^{Opt})}{uRge_i}
	\label{equation7}
\end{gather}

Step 3. $uav_i$ retrieves the $worker_j$ closest to the $task_x^{Opt}$ within its communication range and calculates the time required for $worker_j$ to move to ${tLoc}_x^{Opt}$, see Equation (\ref{equation8}).

\begin{gather}
	\Delta wTimeT_j = \frac{Dis(tLoc_x^{Opt},wLoc_j)}{wRge_j}
	\label{equation8}
\end{gather}

Step 4. $uav_i$ calculates the time required to complete the $task_x^{Opt}$ in collaboration with $worker_j$, denoted as $\Delta taskTime_x^{Opt}$, as well as the remaining online time of $uav_i$, represented as $uLeftT_i$, see Equation (\ref{equation9})(\ref{equation10}).
{\footnotesize {\small \begin{gather}
	\Delta taskTime_x^{Opt} =max(\Delta uTimeT_i,\Delta wTimeT_j) + \frac{T\_costPow_x^{Opt}}{uRge_i}
	\label{equation9}
\end{gather}}}

\begin{gather}
	uLeftT_i = U\_downtime_i-SysTime -\Delta taskTime_x^{Opt}
	\label{equation10}
\end{gather}

Step 5. $uav_i$ calculates the total reduction in its energy incurred by executing $task_x^{Opt}$, denoted as $\Delta uPowSumT_i$, see Equation (\ref{equation11}).
\begin{gather}
	\Delta uPowSumT_i= \Delta uPowT_i+T\_costPow_x^{Opt}
	\label{equation11}
\end{gather}

Step 6. The larger the $uav_i$'s remaining online time ($uLeftT_i$), the more opportunities it has to execute future tasks. Conversely, the greater the total reduction in energy ($\Delta uPowSumT_i$), the higher the cost of $uav_i$ executing $task_x^{Opt}$. Therefore, the expected reward for $uav_i$ should be inversely proportional to both $uLeftT_i$ and $\Delta uPowSumT_i$. In this context, we define the expected reward for $uav_i$ in executing $task_x^{Opt}$ as 1. Accordingly, $uLeftT_i$ and $\Delta uPowSumT_i$ are normalized to the interval (0,1), so that they can be accumulated reasonably. Based on these considerations, we define the expected reward of $uav_i$ for executing $task_x^{Opt}$ as $R\_task_x^{Opt}$, see Equation (\ref{equation12}).
{\small \begin{gather}
	R\_task_x^{Opt} = 1- \frac{uLeftT_i}{U\_downtime_i-U\_uptime_i} \cdot \frac{\Delta uPowSumT_i}{Fullpower_i}
	\label{equation12}
\end{gather}}

\subsubsection*{(2) An estimation of the expected benefit when UAVs are charged:}

Step 1. $uav_i$ calculates the distance between it and the closest charge point ({$charge_y^{Opt} \in Charge$) within its communication range, denoted as $Dis(uLoc_i$,$chLoc_y^{Opt})$
	
Step 2. $uav_i$ calculates the reduction in its remaining energy ($\Delta uPowC_i$) and the required flight time ($\Delta uTimeC_i$) for moving from $uLoc_i$ to $chLoc_y^{Opt}$, see Equation (\ref{equation13})(\ref{equation14}).
\begin{gather}
	\Delta uPowC_i=Dis(uLoc_i,chLoc_y^{Opt})
	\label{equation13}
\end{gather}
\begin{gather}
	\Delta uTimeC_i = \frac{Dis(uLoc_i,chLoc_y^{Opt})}{uRge_i}
	\label{equation14}
\end{gather}

Step 3. $uav_i$ retrieves the $vehicle_k$ closest to the $charge_y^{Opt}$ within its communication range and calculates the time required for $vehicle_k$ to move to $chLoc_y^{Opt}$ ($\Delta vTimeC_k$), see Equation (\ref{equation15}).
\begin{gather}
	\Delta vTimeC_k = \frac{Dis(chLoc_y^{Opt},vLoc_k)}{vRge_k}
	\label{equation15}
\end{gather}

Step 4. $uav_i$ calculates the time required to reach the $charge_y^{Opt}$ and be fully charged by $vehicle_k$, denoted as $\Delta chargeTime_y^{Opt}$, as well as the remaining online time of $uav_i$, represented as $uLeftC_i$, see Equation (\ref{equation16})(\ref{equation17}).
\begin{align}
	\Delta chargeTime_y^{Opt} = max(\Delta uTimeC_i,\Delta vTimeC_k) \notag \\ +\frac{Fullpower_i-(uPow_i-\Delta uPowC_i)}{V\_chargePow_k}
	\label{equation16}
\end{align}
\begin{gather}
	uLeftC_i= U\_downtime_i-SysTime-\Delta chargeTime_v^{Opt}
	\label{equation17}
\end{gather}

Step 5. $uav_i$ calculates the increase in its energy, denoted as $\Delta uPowSumC_i$, see Equation (\ref{equation18}). 
\begin{gather}
	\Delta uPowSumC_i= Fullpower_i-uPow_i
	\label{equation18}
\end{gather}

Step 6. The larger the $uav_i$'s remaining online time ($uLeftC_i$), the more opportunities it has to execute future tasks. The greater the total increase in energy ({$\Delta uPowSumC_i$), the higher the expected benefit of $uav_i$ being fully charged by $vehicle_k$ at $charge_y^{Opt}$. Therefore, the expected reward for $uav_i$ should be proportional to both $uLeftC_i$ and $\Delta uPowSumC_i$. In this context, $uLeftC_i$ and $\Delta uPowSumC_i$ are normalized to the interval (0,1). Based on these considerations, we define the expected reward of $uav_i$ being fully charged by $vehicle_k$ at $charge_y^{Opt}$ as $R\_charge_y^{Opt}$, see Equation (\ref{equation19}).
{\small \begin{align}
	R\_charge_y^{Opt}=\frac{uLeftC_i}{U\_downtime_i-U\_uptime_i} \cdot \frac{\Delta uPowSumC_i}{Fullpower_i}
	\label{equation19}
\end{align}
}

\subsubsection*{(3) Proof of the reasonableness of the expected benefit comparison of UAVs:}

Based on the analysis above, if ${R\_task}_x^{Opt}\geq{R\_charge}_y^{Opt}$, ${uav}_i$ should prefer to execute the task. Otherwise, it should prefer to be charged to extend its endurance.

\textbf{Lemma 2}: The comparison between ${R\_task}_x^{Opt}$ and ${R\_charge}_y^{Opt}$ is reasonable.

The proof of \textbf{Lemma 2} requires addressing the following two aspects: 1) the correlation of ${R\_task}_x^{Opt}$ and ${R\_charge}_y^{Opt}$ with the optimization objective, and 2) the comparability of ${R\_task}_x^{Opt}$ and ${R\_charge}_y^{Opt}$.

1) ${R\_task}_x^{Opt}$ and ${R\_charge}_y^{Opt}$ are positively correlated with the optimization objective ${Cplt\_Tasks\_Per}^{t\times Interval}$ in \textbf{Problem 1}.

As indicated by Equation (12), ${R\_task}_x^{Opt}$ is negatively correlated with ${uLeftT}_i$ and $\Delta uPowSumT_i$.

${uLeftT}_i$ represents the remaining online time of ${uav}_i$ after completing ${task}_x^{Opt}$, and a smaller value of ${uLeftT}_i$ implies that ${uav}_i$ is less likely to continue performing tasks in the future. Thus, the UAV's demand for maintaining a high battery charge is lower, and the demand for executing ${task}_x^{Opt}$ is higher. Therefore, ${uLeftT}_i$ is negatively correlated with ${Cplt\_Tasks\_Per}^{t\times Interval}$.

$\Delta uPowSumT_i$ represents the energy consumed by ${uav}_i$ in executing ${task}_x^{Opt}$, and a smaller value of ${\Delta uPowSumT}_i$ indicates that the cost of executing the task is lower, making the expected benefit from task execution higher. Therefore, ${\Delta uPowSumT}_i$ is negatively correlated with ${Cplt\_Tasks\_Per}^{t\times Interval}$.

By the monotonicity property of the product of function \cite{39-boyd2004convex}, we can conclude that ${R\_task}_x^{Opt}$ is positively correlated with ${Cplt\_Tasks\_Per}^{t\times Interval}$.

As indicated by Equation (19), ${R\_charge}_y^{Opt}$ is positively correlated with ${uLeftC}_i$ and ${\Delta uPowSumC}_i$.

${uLeftC}_i$ represents the remaining online time of ${uav}_i$ after fully charging, and a larger value of ${uLeftC}_i$ implies that ${uav}_i$ is more likely to continue performing tasks in the future. Thus, ${uLeftC}_i$ is positively correlated with ${Cplt\_Tasks\_Per}^{t\times Interval}$.

${\Delta uPowSumC}_i$ represents the increase in endurance after ${uav}_i$ charges, and a larger value of ${\Delta uPowSumC}_i$ means that charging has a greater potential to enhance future task execution. Therefore, ${\Delta uPowSumC}_i$ is positively correlated with ${Cplt\_Tasks\_Per}^{t\times Interval}$.

By the monotonicity property of the product of function, we can conclude that ${R\_charge}_y^{Opt}$ is positively correlated with ${Cplt\_Tasks\_Per}^{t\times Interval}$.

2) The comparability of ${R\_task}_x^{Opt}$ and ${R\_charge}_y^{Opt}$ in terms of their numerical range and physical significance.

Numerical Comparability: From Equation (12), the values of $uLeftT_i/(U\_downtime_i-U\_uptime_i)$ and $\Delta uPowSumT_i/Fullpower_i$ both lie in the range (0,1), which implies that the value of ${R\_task}_x^{Opt}$ also lies in the range (0,1). Similarly, from Equation (19), the values of $uLeftC_i/(U\_downtime_i-U\_uptime_i)$ and $\Delta uPowSumC_i/Fullpower_i$ lie in the range (0,1), so ${R\_charge}_y^{Opt}$ also lies in the range (0,1), which is the same as ${R\_task}_x^{Opt}$.

Physical Comparability: ${R\_task}_x^{Opt}$ reflects the immediate expected benefit of ${uav}_i$ choosing to execute ${task}_x^{Opt}$, while ${R\_charge}_y^{Opt}$ reflects the potential expected benefit of ${uav}_i$ choosing to be charged for future task execution. Both metrics represent expected benefits, making their physical meaning comparable. 

In conclusion, the comparison between ${R\_task}_x^{Opt}$ and ${R\_charge}_y^{Opt}$ is reasonable, thus proving \textbf{Lemma 2}.

\subsection{Heterogeneous Multi-Agent Autonomous Collaborative Scheduling Algorithm}
\label{section 4.2}

We design the framework of heterogeneous multi-agent autonomous cooperative scheduling algorithm to solve the optimal matching results of UAVs, workers and vehicles, as shown in Algorithm \ref{algorithm1}. The algorithm framework is divided into three steps: (1) Self-scheduling behavior decision of the agent ${UWV}_{ijk}(i.e. {uav}_i/{worker}_j/{vehicle}_k)$; (2) Evaluation of the scheduling behavior benefit of the agent ${UWV}_{ijk}$; (3) Local Nash equilibrium determination for the decision-making of the agent ${UWV}_{ijk}$.

\begin{algorithm}[t]
	\caption{Heterogeneous Multi‐Agent Autonomous Collaborative Scheduling}  
	\label{algorithm1}
	\begin{algorithmic}[1]
		\State \textbf{Input:} UAVs $\mathcal{U}$, Workers $\mathcal{W}$, Vehicles $\mathcal{V}$, Tasks $\mathcal{T}$, Chargers $\mathcal{C}$, $LimitTime$, $Interval$.
		\State \textbf{Output:} Schedule $\{A_i^t\}$ for each agent $i$ at each moment $t$.
		\vspace{0.5em}
		\For{$t = 1$ \textbf{to} $(LimitTime/Interval)$}
		\For{each agent $i \in \mathcal{U}\cup\mathcal{W}\cup\mathcal{V}$}  
		\State $\mathcal{P}_i \gets$ set of reachable points within communication range;
		\For{each $n\in\mathcal{P}_i$}
		\State $\mathrm{score}_n \gets -\,\mathrm{Dis}(i,n)$;
		\EndFor
		\State Compute softmax probabilities as Equation (\ref{equation21});
		\State Sample tentative action $A_i^t$ according to $\{p_n\}$;
		\EndFor
		
		\For{each agent $i$}
		\Repeat
		\State $N_i \gets$ set of neighbors of $i$ (within communication range);
		\State Observe $\{A_j^t : j\in N_i\}$;
		\If{$i$ participates in task matching}
		\State $\mathrm{Reward}_i \gets$ number of tasks matched;
		\Else
		\State $\mathrm{Reward}_i \gets$ sum of the power added to all UAVs in $N_i$;
		\EndIf
		\State Compute $S_i$ 
		\Statex\quad
{\footnotesize 		$S_i = \begin{cases}
			1, & \text{if no other action increases }\mathrm{Reward}_i\text{ under fixed }N_i,\\
			0, & \text{otherwise.}
		\end{cases}$}
		\If{$S_i = 0$ \textbf{or} $\exists j\in N_i\text{ s.t. }S_j = 0$}
		\State Resample $A_i^t$ via the softmax step above;
		\EndIf
		\Until{$S_i = 1$ \textbf{and} $\forall j\in N_i: S_j = 1$}
		\State Execute final action $A_i^t$;
		\EndFor
		\EndFor
	\end{algorithmic}
\end{algorithm}

\subsubsection*{(1) Self-scheduling behavior decision of the agent ${UWV}_{ijk}$}

\begin{figure}[b]
	\setlength{\abovecaptionskip}{0.2cm}
	\setlength{\belowcaptionskip}{-0.25cm}
	\centering 
	\includegraphics[width=\linewidth]{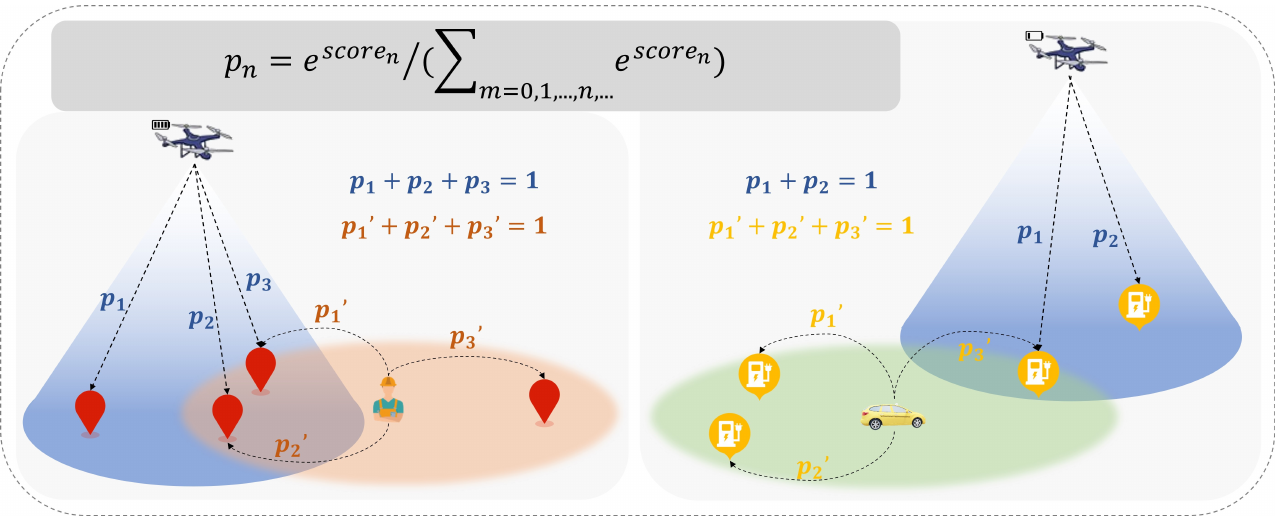}
	\caption{Probabilistic behavior selection of agents.}
	\label{figure4}
\end{figure}

The agent ${UWV}_{ijk}$ commonly adopts a random self-scheduling behavior, which allows it to explore all available behaviors. However, this approach has low exploration efficiency for optimal behaviors, resulting in slower convergence of the algorithm. To improve this, we attempt to assign higher selection probabilities to optimal behaviors, enhancing exploration efficiency and accelerating the algorithm's convergence. Specifically, we use the softmax() function to convert the distances between the agent ${UWV}_{ijk}$ and the potential behavior points into selection probabilities. This method maintains randomness in the behavior selection while ensuring that the agent ${UWV}_{ijk}$ explores directions that lead to faster algorithm convergence, as illustrated in Fig. \ref{figure4}.

The core idea of the softmax() function is to amplify the differences between different input values through exponential expansion, making higher-scoring options more likely to be selected, while ensuring that the sum of the output probabilities is 1. In our problem, the behavior selection probability based on softmax() is calculated as follows: 

a. The agent ${UWV}_{ijk}$ calculates its distance $\{{uwvDis}_0,...,{uwvDis}_n,...\}$ to all selectable behavior points $\{{uwvLoc}_0,...,{uwvLoc}_n,...\}$ within its communication range. The greater the distance (${uwvDis}_n$), the lower the agent ${UWV}_{ijk}$'s gain from moving. Therefore, we set the score for selecting behavior point ${uwvLoc}_n$ as: 
\begin{gather}
	score_n=-uwvDis_n
	\label{equation20}
\end{gather}

b. The probability of selecting behavior point ${uwvLoc}_n$ is calculated as: 
\begin{gather}
	p_n=\frac{e^{score_n}}{\sum\nolimits_{m = 0,...,n,...} {{e^{scor{e_m}}}}}
	\label{equation21}
\end{gather}

Thus, the agent ${UWV}_{ijk}$ selects a self-scheduling behavior within its communication range with probabilities $\{p_0,...,p_n,...\}$. 

\subsubsection*{(2) Evaluation of the scheduling behavior benefit of the agent ${UWV}_{ijk}$}

To guide the agent's self-scheduling behavior in alignment with the optimization goal, we need to design reasonable reward functions for the UAVs, workers and vehicles. 

For the agent ${UWV}_{ijk}$ in the sensing task matching graph, its total reward ${Reward\_task}_{ijk}$ is the number of successful matches between ${uav}_i$, ${worker}_j$ and ${task}_x$ within its communication range: 
\begin{gather}
	matchSign_{ijx} =
	\begin{cases}
		0, & \text{else}\\[1mm]
		1, & \text{if } uLoc_i=wLoc_j=tLoc_x
	\end{cases}
	\label{equation22}
\end{gather}
\begin{gather}
	Reward\_task_{ijk}=\sum matchSign_{ijx}
	\label{equation23}
\end{gather}
where ${matchSign}_{ijx}$ indicates whether the ${uav}_i$, ${worker}_j$ and ${task}_x$ have successfully matched.

For the agent ${UWV}_{ijk}$ in the charging endurance matching graph, its total reward ${Reward\_charge}_{ijk}$ is the sum of the power added to all UAVs within its communication range:
{\small \begin{gather}
	\Delta Pow_{iky} =
	\begin{cases}
		0, & \text{else}\\[1mm]
		Fullpower_i-uPow_i,&\text{if } uLoc_i=vLoc_k=chLoc_y
	\end{cases}
	\label{equation24}
\end{gather}}
\begin{gather}
	Reward\_charge_{ijk}=\sum \Delta Pow_{iky}
	\label{equation25}
\end{gather}
where, ${\Delta Pow}_{iky}$ represents the power added to ${uav}_i$ by ${vehicle}_k$ at ${charge}_y$.

Based on the defined reward functions, the agent ${UWV}_{ijk}$ calculates its current reward indicator $S_{ijk}$, which indicates whether it can increase its reward value ${Reward\_task}_{ijk}$ or ${Reward\_charge}_{ijk}$ by merely adjusting its self-scheduling behavior. If so, the agent's reward indicator $S_{ijk}$ is set to 0; otherwise, it is set to 1.

\subsubsection*{(3) Local Nash equilibrium determination for the decision-making of the agent ${UWV}_{ijk}$}

To ensure the overall stability of the scheduling system, the agent must ensure that its self-scheduling behavior leads to a local Nash equilibrium. 

The agent ${UWV}_{ijk}$ can obtain the reward indicators of other agents within its communication range. If the reward indicators of all other agents in its range are 1, the agent ${UWV}_{ijk}$ is considered to have reached a local nash equilibrium and will execute its self-scheduling behavior decision. Otherwise, the agent will not execute its self-scheduling behavior decision and will reselect its self-scheduling behavior.

\textbf{Lemma 3}: When the coupling strength $\varepsilon$ between multi-agent communication covers is low, the local nash equilibrium solution approximates the global nash equilibrium solution.

To characterize the degree of mutual influence in multi-agent decision-making, we define the coupling strength $\varepsilon$. Assume that there are N agents $\{...,UWV_{ijk},...,{UWV_{ijk+n},...\}}$ in the system, each with a defined communication range $\{...,Cover_{ijk},...,{Cover_{ijk+n},...\}}$. Then, the coupling strength $\varepsilon$ is defined as:

\begin{gather}
	\varepsilon=\frac{\sum_{ijk=1}^{N}\sum_{ijk+n}^{N}(\frac{Cover_{ijk} \cap Cover_{ijk+n}}{Cover_{ijk} \cup Cover_{ijk+n}})}{C(N,2)}
	\label{equation26}
\end{gather}}

where, ${Cover}_{ijk}\cap{Cover}_{ijk+n}$ is the overlapping area of the communication ranges of ${UWV}_{ijk}$ and ${UWV}_{ijk+n}$, ${Cover}_{ijk}\cup{Cover}_{ijk+n}$ is their union, and $C(N,2)$ is the combination number of N agents taken 2 at a time. 

When all agents are located at the same position, $\varepsilon$ equals 1, and the coupling strength among the agents is maximized; Conversely, if there is no overlap in the communication ranges of the agents, $\varepsilon$ equals 0, resulting in the minimum coupling strength. As illustrated in Fig. \ref{figure5}, for example, ${worker}_3$’s communication range overlaps with those of ${worker}_2$, ${worker}_4$ and ${uav}_2$. Therefore, when determining whether to execute self-scheduling behavior, ${worker}_3$ must comprehensively consider the reward indices from all three, indicating a strong mutual influence on decision making among the agents. In contrast, since the communication range of ${uav}_5$ is relatively isolated, its decision regarding self-scheduling is based solely on its own reward index without being affected by the decisions of other agents.

\begin{figure}[ht]
	\setlength{\abovecaptionskip}{0.2cm}
	\setlength{\belowcaptionskip}{-0.25cm}
	\centering 
	\includegraphics[width=\linewidth]{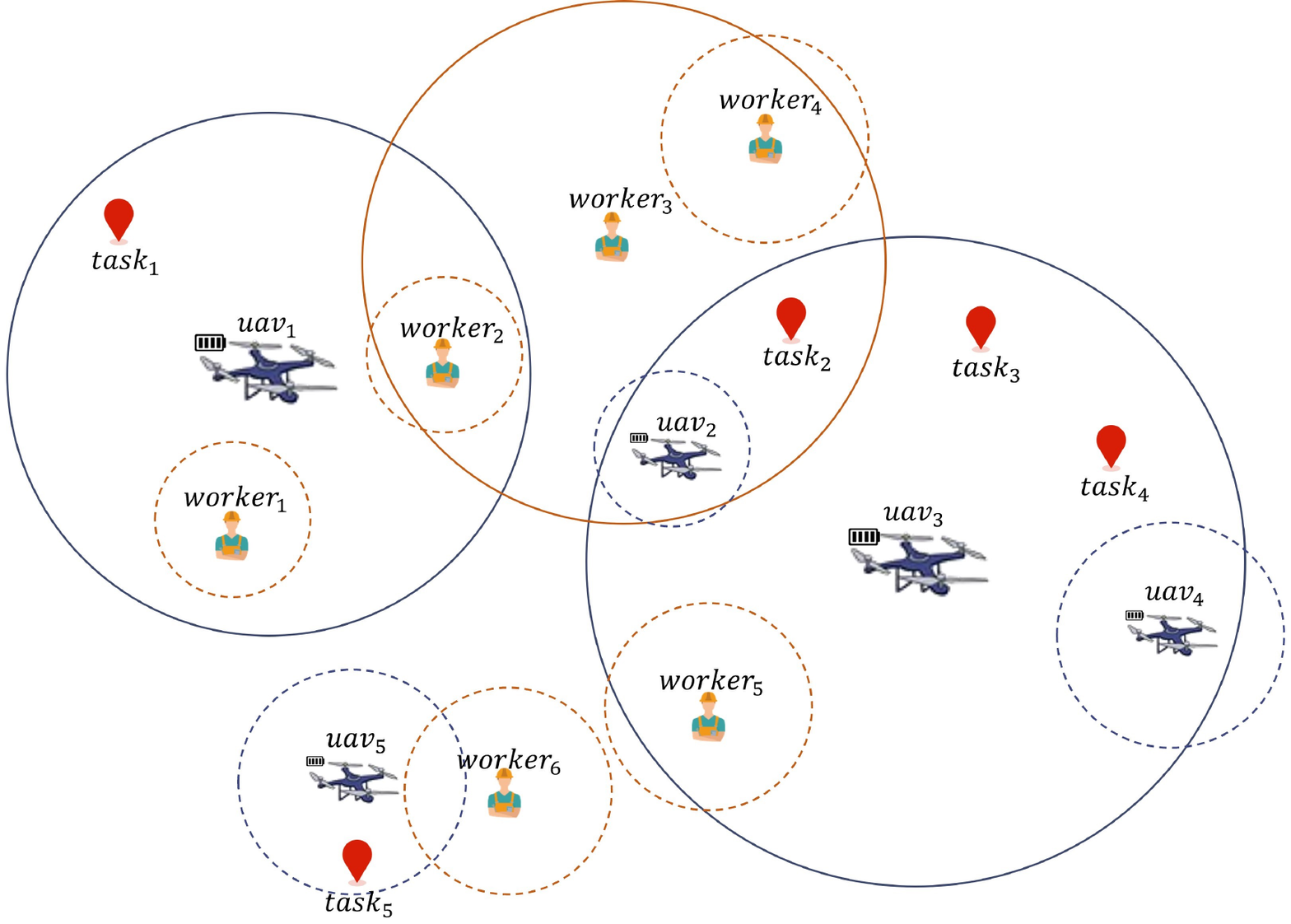}
	\caption{Schematic diagram of sensing task matching process.}
	\label{figure5}
\end{figure}

Proof of \textbf{Lemma 3} for the Sensing Task Matching Process:

We aim to find a constant $\alpha$ such that:
\begin{gather}
	\sum{Reward\_task_{ijk}}=\alpha \cdot Cplt\_Tasks\_Per^{t \times Interval}
	\label{equation27}
\end{gather}
Assume that the influence among the multi-agent behavioral strategies is governed by a coupling strength $\varepsilon$. That is, there exists a constant $C$ such that
\begin{gather}
	Cplt\_Tasks\_Per^{t \times Interval}-\sum{Reward\_task_{ijk}}= C \cdot \varepsilon
	\label{equation28}
\end{gather}
Hence, we obtain 
\begin{gather}
	\sum{Reward\_task_{ijk}}=Cplt\_Tasks\_Per^{t \times Interval}-C \cdot \varepsilon
	\label{equation29}
\end{gather}
After normalizing, we get 
{\small \begin{align}
	\frac{\sum{Reward\_task_{ijk}}}{Cplt\_Tasks\_Per^{t \times Interval}} = 1- \frac{C \cdot \varepsilon}{Cplt\_Tasks\_Per^{t \times Interval}}
	\label{equation30}
\end{align}}
We define the approximation factor as 
\begin{gather}
	\alpha = 1-\frac{C \cdot \varepsilon}{Cplt\_Tasks\_Per^{t \times Interval}}
	\label{equation31}
\end{gather}

In \textbf{Problem 1}, since the numbers of agents (i.e. UAVs, workers, and vehicles) are finite and the communication coverage range of each agent is fixed, the coupling strength $\varepsilon$ in the multi-agent communication network is bounded. As $\varepsilon \to 0$, it follows that $\alpha \to 1$, which implies that $\sum{Reward\_task}_{ijk}$ can closely approximate ${Cplt\_Tasks\_Per}^{t\times Interval}$. This completes the proof of \textbf{Lemma 3}. 

A similar proof applies for \textbf{Lemma 3} based on the charging endurance matching process.

\section{Experiment}
\label{section 5}

\subsection{Dataset Setup}
\label{section 5.1}
In this work, we comprehensively evaluate the performance of our proposed algorithm based on both real-world datasets and simulated datasets. The detailed descriptions of the datasets are as follows.

\subsubsection*{(1) Real-world datasets}
The real-world datasets are derived from practical scenarios, and they are capable of reflecting the task distribution, agent behavior patterns, and resource constraints in actual environments. The real-world datasets include BicycleOrder data \cite{40-li2015traffic} \cite{41-zheng2014urban}, TaxiTrajectories data \cite{42-yu2021object}, and DidiOrders data \cite{43-didi}. First, the coverage area of the real-world datasets is discretized. Subsequently, the data is filtered, classified, and annotated to obtain datasets that conform to the experimental specifications; the specific processing procedures can be found in \cite{github}.
\subsubsection*{(2) Simulated datasets}
The randomly simulated datasets are generated through a parameterized process, which allows for flexible adjustments of the area scale, the number of tasks, the number of agents, and so forth, thus facilitating the validation of the algorithm under various scenarios. The introduction of simulated data compensates for the limitations in diversity and controllability found in the real-world datasets. Detailed data generation procedures can be found in \cite{github}.

\begin{figure*}[ht]
	\setlength{\abovecaptionskip}{0.2cm}
	\setlength{\belowcaptionskip}{-0.25cm}
	\centering 
	\includegraphics[width=\linewidth]{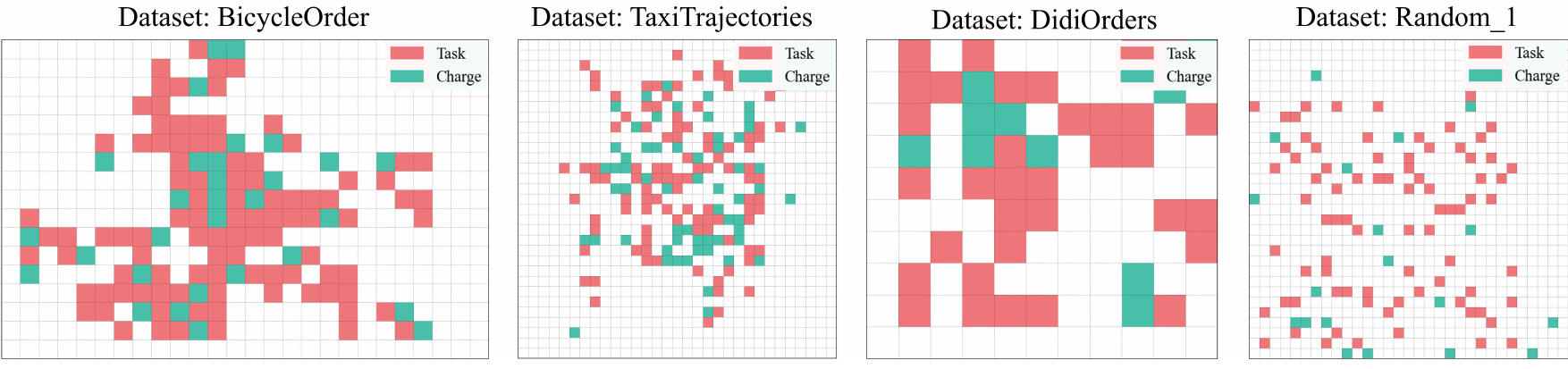}
	\caption{The distribution of task points and charges for the four types of datasets (each grid cell is 1 km * 1 km).}
	\label{figure6}
\end{figure*}

Specifically, the distribution patterns of the task points and charges in the four datasets vary significantly, as illustrated in Fig. \ref{figure6}. A detailed description of the datasets is provided in Table \ref{table1}.

\begin{table*}[htbp]
	\centering
	\footnotesize
	\caption{Experimental dataset description.}
	\label{table1}
	\renewcommand{\arraystretch}{1.5}
	\resizebox{\textwidth}{!}{%
		\begin{tabular}{c|ccccccc}
			\hline
			Data & Area number & Tasks number & Charges number & Agents online time & Agent number & Task cost & Charging power \\
			\hline
			BicycleOrder & $25\times16$ & 99 & 30 & $>70$min & (54,17,34) & 3 & 10 \\
			\hline
			TaxiTrajectories & $30\times30$ & 102 & 60 & $>90$min & (76,32,40) & 3 & 10 \\
			\hline
			DidiOrders   & $10\times9$  & 34 & 10 & $>50$min & (34,18,24) & 3 & 10 \\
			\hline
			Random\_1   & $30\times30$ & 80 & 20 & 60min & (50,30,20) & 3 & 10 \\
			\hline
			Random\_2   & $20\times20$ & 80 & 20 & 60min & (50,30,20) & 3 & 10 \\
			\hline
			Random\_3   & $40\times40$ & 80 & 20 & 60min & (50,30,20) & 3 & 10 \\
			\hline
			Random\_4   & $30\times30$ & 60 & 20 & 60min & (50,30,20) & 3 & 10 \\
			\hline
			Random\_5   & $30\times30$ & 100 & 20 & 60min & (50,30,20) & 3 & 10 \\
			\hline
			Random\_6   & $30\times30$ & 80 & 15 & 60min & (50,30,20) & 3 & 10 \\
			\hline
			Random\_7   & $30\times30$ & 80 & 25 & 60min & (50,30,20) & 3 & 10 \\
			\hline
			Random\_8   & $30\times30$ & 80 & 20 & 40min & (50,30,20) & 3 & 10 \\
			\hline
			Random\_9   & $30\times30$ & 80 & 20 & 80min & (50,30,20) & 3 & 10 \\
			\hline
			Random\_10  & $30\times30$ & 80 & 20 & 60min & (30,20,10) & 3 & 10 \\
			\hline
			Random\_11  & $30\times30$ & 80 & 20 & 60min & (70,40,30) & 3 & 10 \\
			\hline
			Random\_12  & $30\times30$ & 80 & 20 & 60min & (30,30,20) & 3 & 10 \\
			\hline
			Random\_13  & $30\times30$ & 80 & 20 & 60min & (70,30,20) & 3 & 10 \\
			\hline
			Random\_14  & $30\times30$ & 80 & 20 & 60min & (50,20,20) & 3 & 10 \\
			\hline
			Random\_15  & $30\times30$ & 80 & 20 & 60min & (50,40,20) & 3 & 10 \\
			\hline
			Random\_16  & $30\times30$ & 80 & 20 & 60min & (50,30,10) & 3 & 10 \\
			\hline
			Random\_17  & $30\times30$ & 80 & 20 & 60min & (50,30,30) & 3 & 10 \\
			\hline
			Random\_18  & $30\times30$ & 80 & 20 & 60min & (50,30,20) & 2 & 10 \\
			\hline
			Random\_19  & $30\times30$ & 80 & 20 & 60min & (50,30,20) & 4 & 10 \\
			\hline
			Random\_20  & $30\times30$ & 80 & 20 & 60min & (50,30,20) & [2,3] & 10 \\
			\hline
			Random\_21  & $30\times30$ & 80 & 20 & 60min & (50,30,20) & [3,4] & 10 \\
			\hline
			Random\_22  & $30\times30$ & 80 & 20 & 60min & (50,30,20) & [4,5] & 10 \\
			\hline
			Random\_23  & $30\times30$ & 80 & 20 & 60min & (50,30,20) & 3 & 8 \\
			\hline
			Random\_24  & $30\times30$ & 80 & 20 & 60min & (50,30,20) & 3 & 12 \\
			\hline
			Random\_25  & $30\times30$ & 80 & 20 & 60min & (50,30,20) & 3 & [6,8] \\
			\hline
			Random\_26  & $30\times30$ & 80 & 20 & 60min & (50,30,20) & 3 & [8,10] \\
			\hline
			Random\_27  & $30\times30$ & 80 & 20 & 60min & (50,30,20) & 3 & [10,12] \\
			\hline
		\end{tabular}
	}
\end{table*}

\subsection{Experiment Setup}
\label{section 5.2}

Based on the above data set, we set up six groups of experiments to comprehensively evaluate the HoAs-PALN algorithm, see Table \ref{table2}. All experiments were carried out on a dedicated workstation with the following configuration: an \mbox{Intel\textregistered{} Core\texttrademark{} i7-10700 CPU} running at a base frequency of 2.90\,GHz, 16\,GB of RAM. The host operating system was Microsoft Windows~11 (64-bit, version~23H2), and the software environment was built around Python~3.9.0.

\begin{table*}[htbp]
	\centering
	\footnotesize 
	\caption{Comparison of experimental group settings.}
	\label{table2}        
	\renewcommand{\arraystretch}{1.5}    
	\setlength{\tabcolsep}{2mm}
	\begin{tabularx}{\textwidth}{c|*{5}{Y}}
		\hline
		Group 
		& Data 
		& Radius 
		& Interval 
		& LimitTime 
		& Control variable \\
		\hline
		(1) 
		& \makecell{BicycleOrder\\TaxiTrajectories\\DidiOrders\\Random\_1} 
		& 8 
		& 5min/10min/15min 
		& 3h 
		& $Interval$ \\
		\hline
		(2) 
		& \makecell{BicycleOrder\\TaxiTrajectorie\\DidiOrders\\Random\_1} 
		& 8 
		& 5min 
		& 2h/3h/4h 
		& $LimitTime$ \\
		\hline
		(3) 
		& \makecell{Random\_1,2,3\\Random\_1,4,5\\Random\_1,6,7\\Random\_1,8,9} 
		& 8 
		& 5min 
		& 3h 
		& \makecell{Area number/\\Tasks number/\\Charges number/\\Agents online time} \\
		\hline
		(4) 
		& \makecell{Random\_1,10,11\\Random\_1,12,13\\Random\_1,14,15\\Random\_1,16,17} 
		& 8 
		& 5min 
		& 3h 
		& \makecell{Agents number/\\Workers number/\\UAVs number/\\Vehicles number} \\
		\hline
		(5) 
		& \makecell{Random\_1,18,19\\Random\_20,21,22\\Random\_1,23,24\\Random\_25,26,27} 
		& 8 
		& 5min 
		& 3h 
		& \makecell{Task cost/\\Charging power} \\
		\hline
		(6) 
		& \makecell{BicycleOrder\\TaxiTrajectories\\DidiOrders\\Random\_1} 
		& 6/8/10 
		& 5min 
		& 3h 
		& $Radius$ \\
		\hline
	\end{tabularx}
\end{table*}

\subsection{Comparison algorithm}
\label{section 5.3}

\subsubsection*{(1) GREEDY}
The greedy algorithm makes real-time decisions based on a local optimal strategy \cite{44-li2019three}\cite{45-wang2022task}. Each agent (UAV, worker, vehicle) independently selects the action that yields the highest immediate benefit: the UAV chooses the nearest task point or charge point after dimensionality reduction partitioning, the worker selects the closest task point, and the vehicle selects the closest charge point. In cases where multiple agents of the same type compete for the same task, the match with the minimum distance is retained. A sensing task is considered successfully matched only if it is simultaneously assigned to both a UAV and a worker; similarly, a charge point must be concurrently matched to both a UAV and an vehicle for the match to be valid.

\subsubsection*{(2) K-WTA (K -Winners-Take-All)}
Originating from the selective activation mechanism in neural networks, K-WTA can be utilized for task matching among multiple agents \cite{46-liu2024distributed}\cite{47-liu2024distributed}\cite{48-qi2021robust}. In this approach, the UAV is divided into task UAVs and charging UAVs based on dimensionality reduction partitioning. Taking task matching as an example, the task UAVs and workers assess the task points within their communication ranges (e.g. based on distance), retain the top ${K_1}$ and ${K_2}$ highest-scoring tasks, and then identify the intersection of these sets to determine the triadic match for collaborative execution. If no intersection exists, tasks are sequentially attempted in descending order of score. Matching for charging follows a similar procedure.

\subsubsection*{(3) MADL (Multi-Agent Deep Learning)}
Based on a multi-agent deep learning framework \cite{liu2019energy} \cite{liu2020energy}, MADL employs a shared neural network to model the state and action spaces of heterogeneous agents. Global information (such as the distribution of tasks and charges) is processed through a convolutional network for feature extraction, which is then combined with local state information (e.g. agent positions and UAV battery levels) as input to a deep network. This network outputs the instantaneous action values for each agent and is optimized end-to-end to maximize the immediate task completion rate.

\subsubsection*{(4) MARL (Multi-Agent Reinforcement Learning)}
Multi-agent reinforcement learning optimizes the long-term cumulative reward via cooperative strategies \cite{ye2023exploring} \cite{zhao2022cadre} \cite{wang2022human}. Based on the QMIX algorithm, a mixing network is designed to integrate the individual local Q-values of all agents, with non-negative weight constraints to preserve monotonicity. Agents balance immediate benefits and future cooperative potential through an exploration–exploitation strategy. The reward function is focused on the number of tasks completed, thereby avoiding short-term interference from charging/endurance gains. To mitigate training instability, a target network is introduced to support complex spatiotemporal cooperative decision-making. Note: the detailed implementations of both MADL and MARL closely refer to \cite{10-han2024collaborative}.

\subsubsection*{(5) HoAs-RALN (Heterogeneous Multi-Agent Online Autonomous Collaborative Scheduling Algorithm Based on Random Action Selection and Local Nash Game)}
Unlike the method in which agents convert their distances to optional behavior points into selection probabilities through the softmax() function and make matching selections based on those probabilities, this method employs an unbiased pure random strategy for distributed decision-making. Each agent (UAV, worker, vehicle) randomly selects actions within the communication range: the task UAV randomly selects a task point, the worker randomly selects a task point, and collaboration is triggered only when both select the same location. The matching for charging follows the same principle.

\subsubsection*{(6) LLM (Large Language Model)}
Capitalizing on the strong generalization and real-time reasoning capabilities of large language models in solving combinatorial optimization problems, we introduce an online decision-making method based on LLMs \cite{49-li2024urbangpt}\cite{50-yuan2024unist}. In this approach, the problem description is translated into a natural language input, and the large model generates a scheduling plan in real time. For our problem, detailed natural language prompts sent to the large model can be found in \cite{github}. After extensive experiments and interactions with several state-of-the-art large models, the results consistently indicated that the current LLMs’ comprehension and reasoning abilities are insufficient for our complex post-disaster environmental sensing scheduling problem. They frequently exhibit unreasonable outcomes or erroneous inferences in understanding and predicting scheduling behaviors, failing to fully consider the various constraints inherent in our research context. 

\subsection{Experimental Results}
\label{section 5.4}
In order to control variables and accurately evaluate the impact of different factors on algorithm performance, all parameters are set to default values except for the specific variables adjusted in each experimental group. The default settings are: area number is 30×30, tasks number is 80, charges number is 20, agents online time is 60 minutes, workers number is 50, UAVs number is 30, vehicles number is 20, a task cost is 3, and a vehicle’s charging power is 10. In each experiment, only the target variable is modified, while all other parameters remain at their default values.

\subsubsection*{(1) Experimental results with different decision $Interval$}
\begin{figure}[h]
	\setlength{\abovecaptionskip}{0.2cm}
	\setlength{\belowcaptionskip}{-0.25cm}
	\centering
	\includegraphics[width=\linewidth]{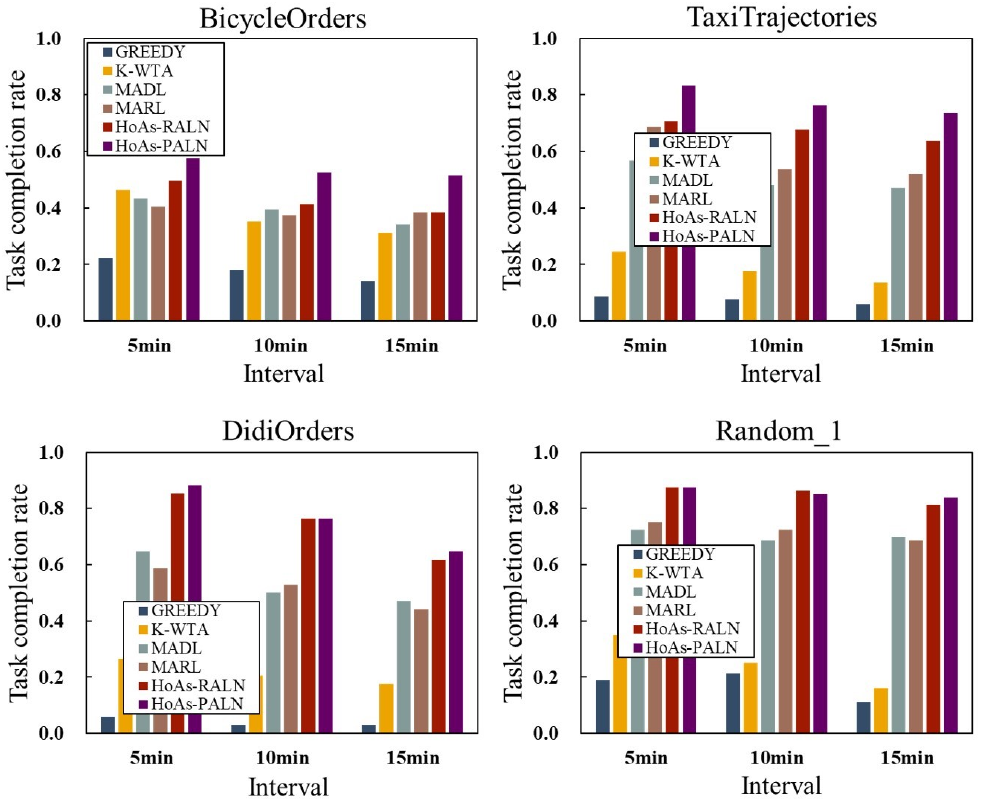}
	\caption{Experimental results with different $Interval$.}
	\label{figure7}
\end{figure}

As the $Interval$ increases from 5 to 15 minutes, HoAs-PALN consistently performs well, see Fig. \ref{figure7}. Specifically, HoAs-PALN achieves task completion rates of 79.16\%, 72.62\%, and 68.38\% at 5, 10, and 15 minutes, representing improvements of 65.25\%, 55.83\%, and 59.83\% over GREEDY; 46.05\%, 47.97\%, and 48.65\% over K-WTA; 19.79\%, 21.08\%, and 18.77\% over MADL; and 18.45\%, 18.44\%, and 17.58\% over MARL, respectively. These results demonstrate that, although completion rates decline slightly as the $Interval$ lengthens, HoAs-PALN maintains an advantage under all different $Interval$.

\subsubsection*{(2) Experimental results with different $LimitTime$}
\begin{figure}[h]
	\setlength{\abovecaptionskip}{0.2cm}
	\setlength{\belowcaptionskip}{-0.25cm}
	\centering 
	\includegraphics[width=\linewidth]{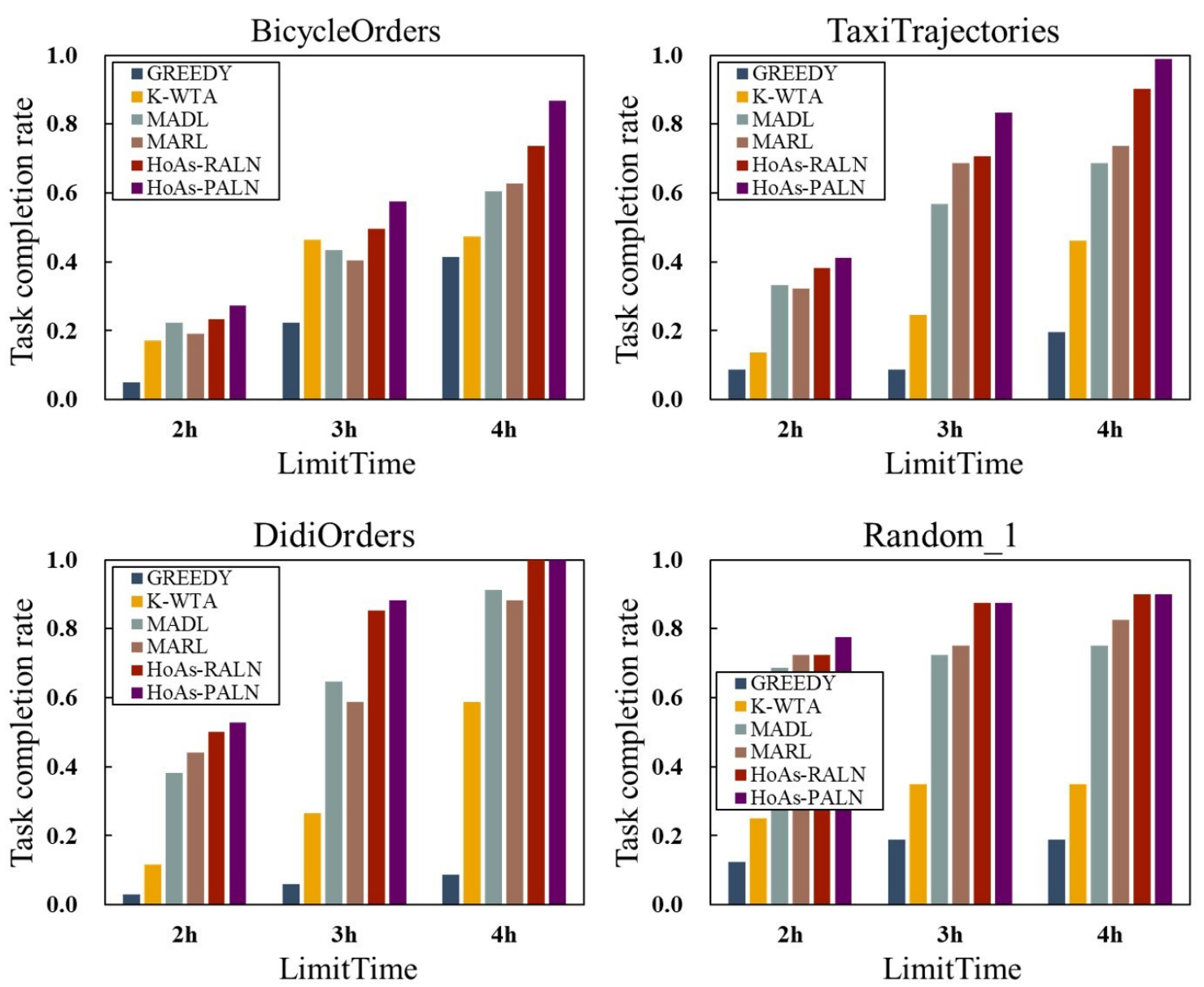}
	\caption{Experimental results with different $LimitTime$.}
	\label{figure8}
\end{figure}

As the $LimitTime$ increases from 2 to 4 hours, HoAs-PALN consistently outperforms all baselines, see Fig. \ref{figure8}. Specifically, HoAs-PALN achieves task completion rates of 49.72\%, 79.16\%, and 93.97\% at 2, 3, and 4 hours, representing improvements of 42.39\%, 65.24\%, and 71.82\% over GREEDY; 32.81\%, 46.05\%, and 47.13\% over K-WTA; 9.09\%, 19.79\%, and 20.12\% over MADL; and 7.68\%, 18.45\%, and 17.25\% over MARL, respectively. These findings indicate that HoAs-PALN’s optimization potential is increasingly realized in longer-duration environments.

\subsubsection*{(3) Experimental results with different area number, tasks number, charges number, and agents online time}
\begin{figure}[h]
	\setlength{\abovecaptionskip}{0.2cm}
	\setlength{\belowcaptionskip}{-0.25cm}
	\centering 
	\includegraphics[width=\linewidth]{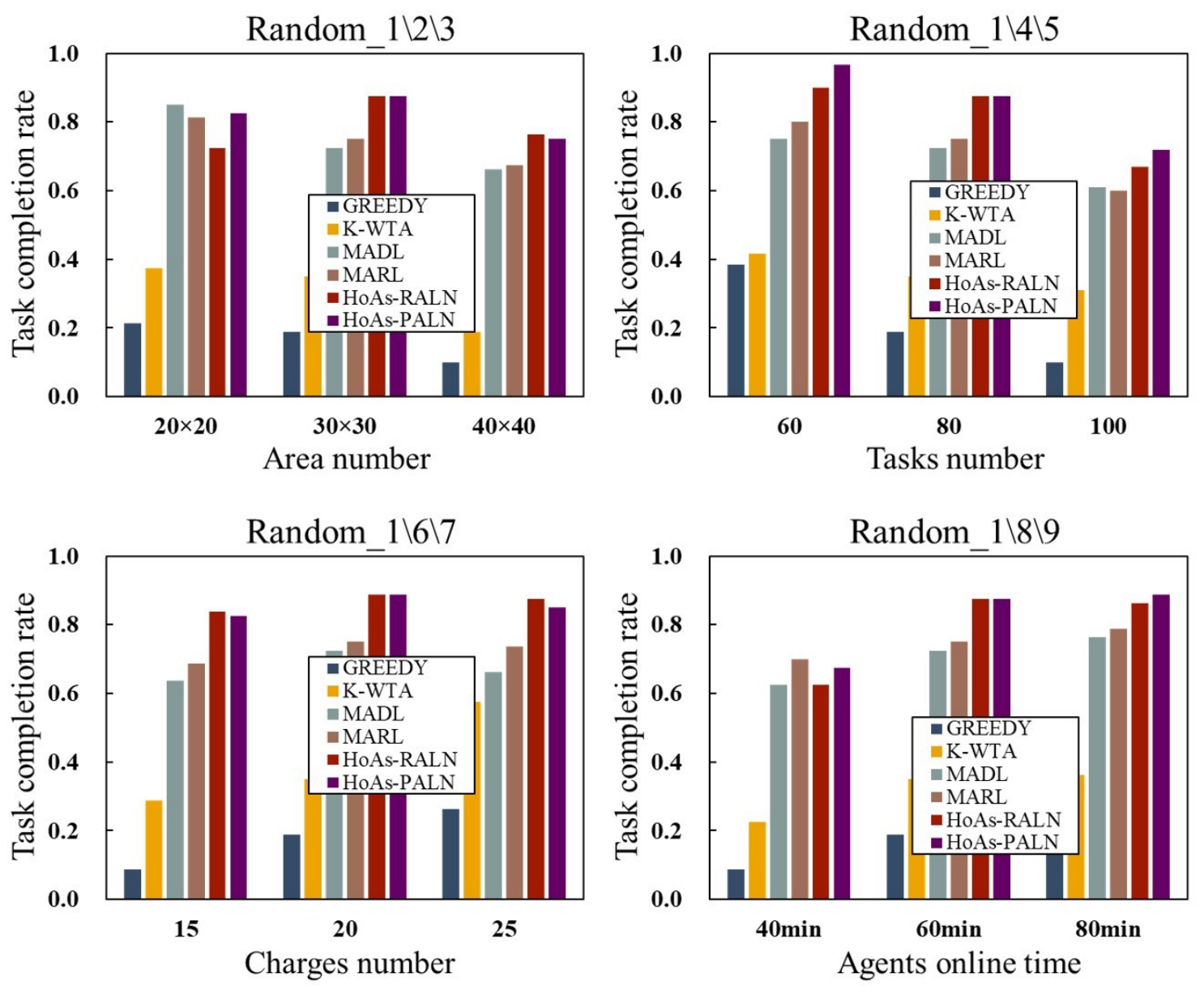}
	\caption{Experimental results with different area number, tasks number, charges number, and agents online time.}
	\label{figure9}
\end{figure}

According to Fig. \ref{figure9}, the following information can be obtained. As the area number expands from 20×20 to 40×40, HoAs‑PALN records task‑completion rates of 82.50 \%, 87.50 \%, and 75.00 \%, representing advantages of 61.25\%, 68.75\%, and 65.00\% over GREEDY; 45.00\%, 52.50\%, and 56.25\% over K‑WTA; –2.50\%(slightly decreased), 15.00\%, and 8.75\% over MADL; and 1.25\%, 12.50\%, and 7.50\% over MARL, respectively. As the tasks number rises from 60 to 100, its rates decline from 96.67\% through 87.50\% to 72.00\%, yet still exceed GREEDY by 58.34\%, 68.75\%, 62.00\%; K‑WTA by 55.00\%, 52.50\%, 41.00\%; MADL by 21.67\%, 15.00\%, 11.00\%; and MARL by 16.67\%, 12.50\%, 12.00\%. Increasing the charges number from 15 to 25 yields completion rates of 82.50\%, 87.50\%, 85.00\%, surpassing GREEDY by 73.75\%, 68.75\%, 58.75\%; K‑WTA by 53.75\%, 52.50\%, 27.50\%; MADL by 18.75\%, 15.00\%, 18.75\%; and MARL by 13.75\%, 12.50\%, 11.25\%. Finally, extending agent online time from 40 to 80 minutes lifts HoAs‑PALN from 67.50\% to 88.75\%, corresponding to gains of 58.75\%, 68.75\%, 70.00\% over GREEDY; 45.00\%, 52.50\%, 52.50\% over K‑WTA; 5.00\%, 15.00\%, 12.50\% over MADL; and –2.50\%(slightly decreased), 12.50\%, 10.00\% over MARL. These comprehensive results indicate that despite absolute performance degradation in more challenging environments, HoAs PALN consistently maintains good performance.

\subsubsection*{(4) Experimental results with different agents number}
\begin{figure}[ht]
	\setlength{\abovecaptionskip}{0.2cm}
	\setlength{\belowcaptionskip}{-0.25cm}
	\centering 
	\includegraphics[width=\linewidth]{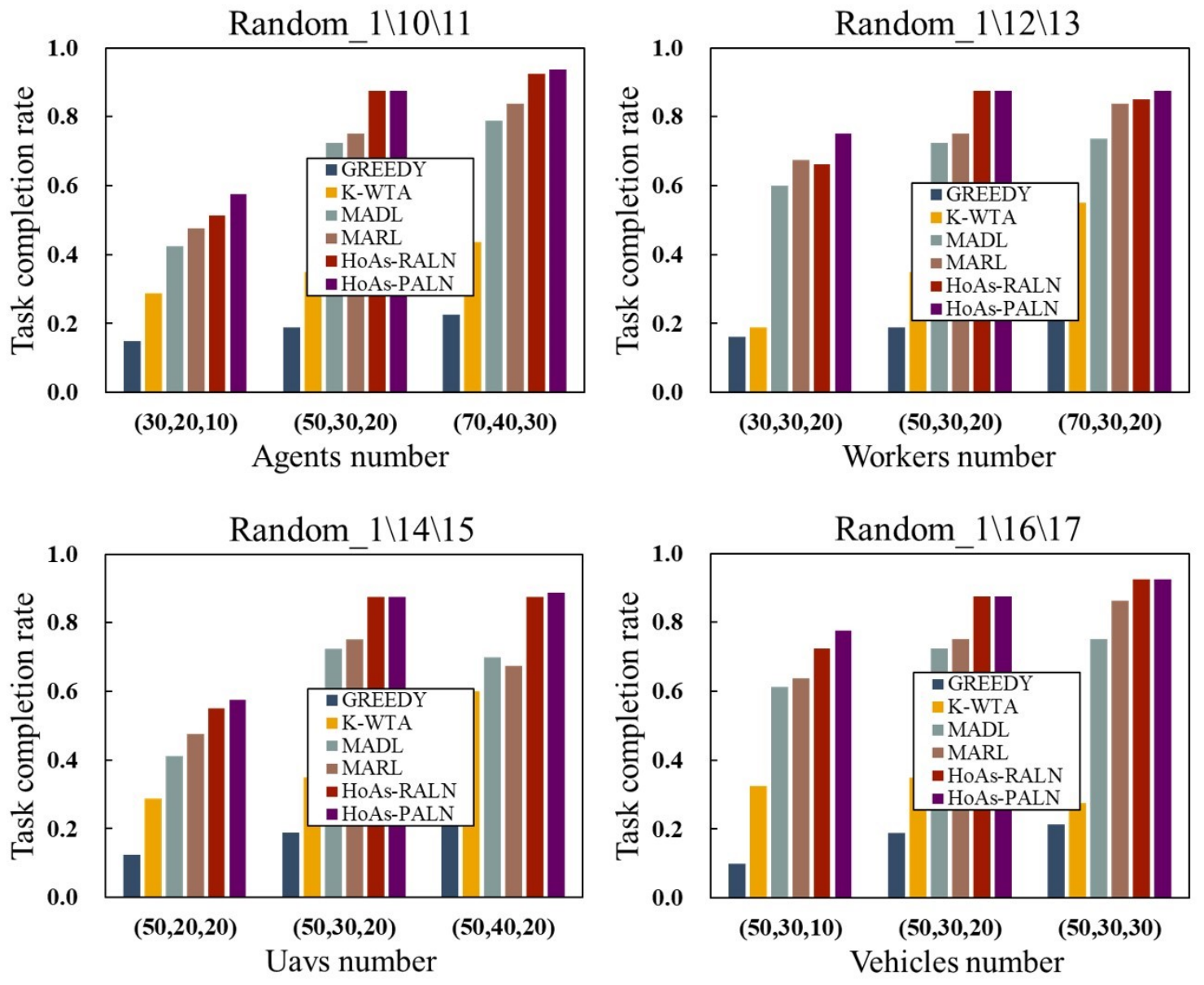}
	\caption{Experimental results with different agents number.}
	\label{figure10}
\end{figure}

According to Fig. \ref{figure10}, the following information can be obtained. As the agents number expands from \((30,20,10)\) to \((50,30,20)\) to \((70,50,30)\), HoAs-PALN records task completion rates of 57.50\%, 87.50\%, and 93.75\%, representing advantages of 42.50\%, 68.75\%, and 71.25\% over GREEDY; 28.75\%, 52.50\%, and 50.00\% over K-WTA; 15.00\%, 15.00\%, and 15.00\% over MADL; and 10.00\%, 12.50\%, and 10.00\% over MARL, respectively. As the workers number rises from 30 to 50 to 70 (with UAVs and vehicles fixed at 30 and 20), its rates climb from 75.00\% through 87.50\% to 87.50\%, yielding gains of 58.75\%, 68.75\%, and 41.25\% over GREEDY; 56.25\%, 52.50\%, and 32.50\% over K-WTA; 15.00\%, 15.00\%, and 13.75\% over MADL; and 7.50\%, 12.50\%, and 3.75\% over MARL. As the UAVs number expands from 20 to 30 to 40 (with workers and vehicles at 50 and 20), completion rates rise from 57.50\% through 87.50\% to 88.75\%, corresponding to improvements of 45.00\%, 68.75\%, and 58.75\% over GREEDY; 28.75\%, 52.50\%, and 28.75\% over K-WTA; 16.25\%, 15.00\%, and 18.75\% over MADL; and 10.00\%, 12.50\%, and 21.25\% over MARL. As the vehicles number increases from 10 to 20 to 30 (with workers and UAVs at 50 and 30), its completion rates grow from 77.50\% through 87.50\% to 92.50\%, representing advantages of 67.50\%, 68.75\%, and 71.25\% over GREEDY; 45.00\%, 52.50\%, and 65.00\% over K-WTA; 16.25\%, 15.00\%, and 17.50\% over MADL; and 13.75\%, 12.50\%, and 6.25\% over MARL. These results suggest that as the number of agents increases, the task completion rates of all methods improve.

\subsubsection*{(5) Experimental results with different task cost and charging power}
\begin{figure}[ht]
	\setlength{\abovecaptionskip}{0.2cm}
	\setlength{\belowcaptionskip}{-0.25cm}
	\centering 
	\includegraphics[width=\linewidth]{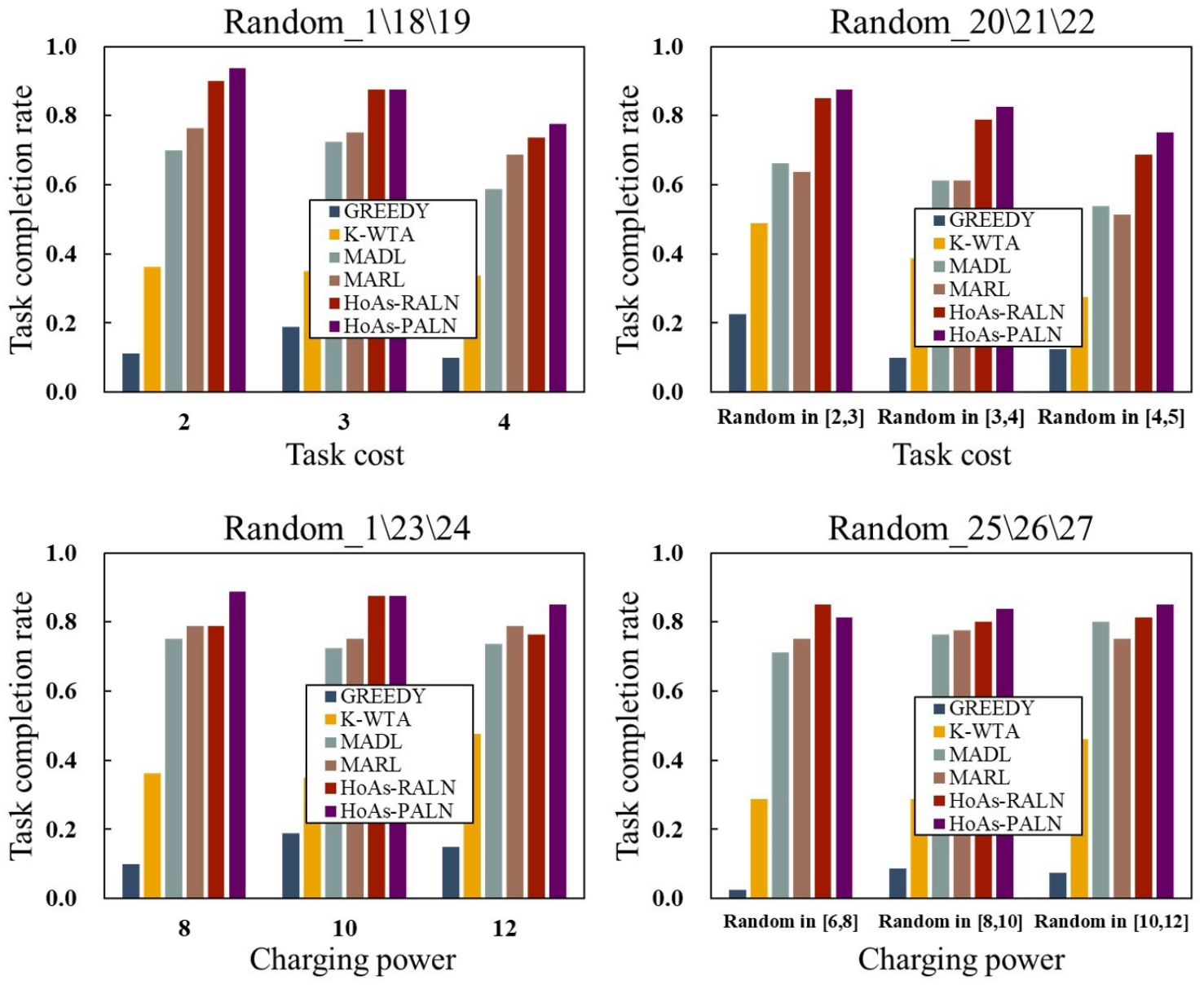}
	\caption{Experimental results with different task cost and charging power.}
	\label{figure11}
\end{figure}

According to Fig. \ref{figure11}, the following information can be obtained.As task cost rises from 2 to 3 to 4, and when it varies over [2,3], [3,4] and [4,5], HoAs-PALN achieves task completion rates of 93.75\%, 87.50\%, 77.50\%, 87.50\%, 82.50\% and 75.00\%, corresponding to gains of 82.50\%, 68.75\%, 67.50\%, 65.00\%, 72.50\% and 62.50\% over GREEDY; 57.50\%, 52.50\%, 43.75\%, 38.75\%, 43.75\% and 47.50\% over K-WTA; 23.75\%, 15.00\%, 18.75\%, 21.25\%, 21.25\% and 21.25\% over MADL; and 17.50\%, 12.50\%, 8.75\%, 23.75\%, 21.25\% and 23.75\% over MARL. As vehicle charging power increases from fixed values of 8, 10 and 12 and ranges [6,8], [8,10] and [10,12], HoAs-PALN records completion rates of 88.75\%, 87.50\%, 85.00\%, 81.25\%, 83.75\% and 85.00\%, yielding advantages of 78.75\%.

\subsubsection*{(6) Experimental results with different $Radius$}
\begin{figure}[ht]
	\setlength{\abovecaptionskip}{0.2cm}
	\setlength{\belowcaptionskip}{-0.25cm}
	\centering 
	\includegraphics[width=\linewidth]{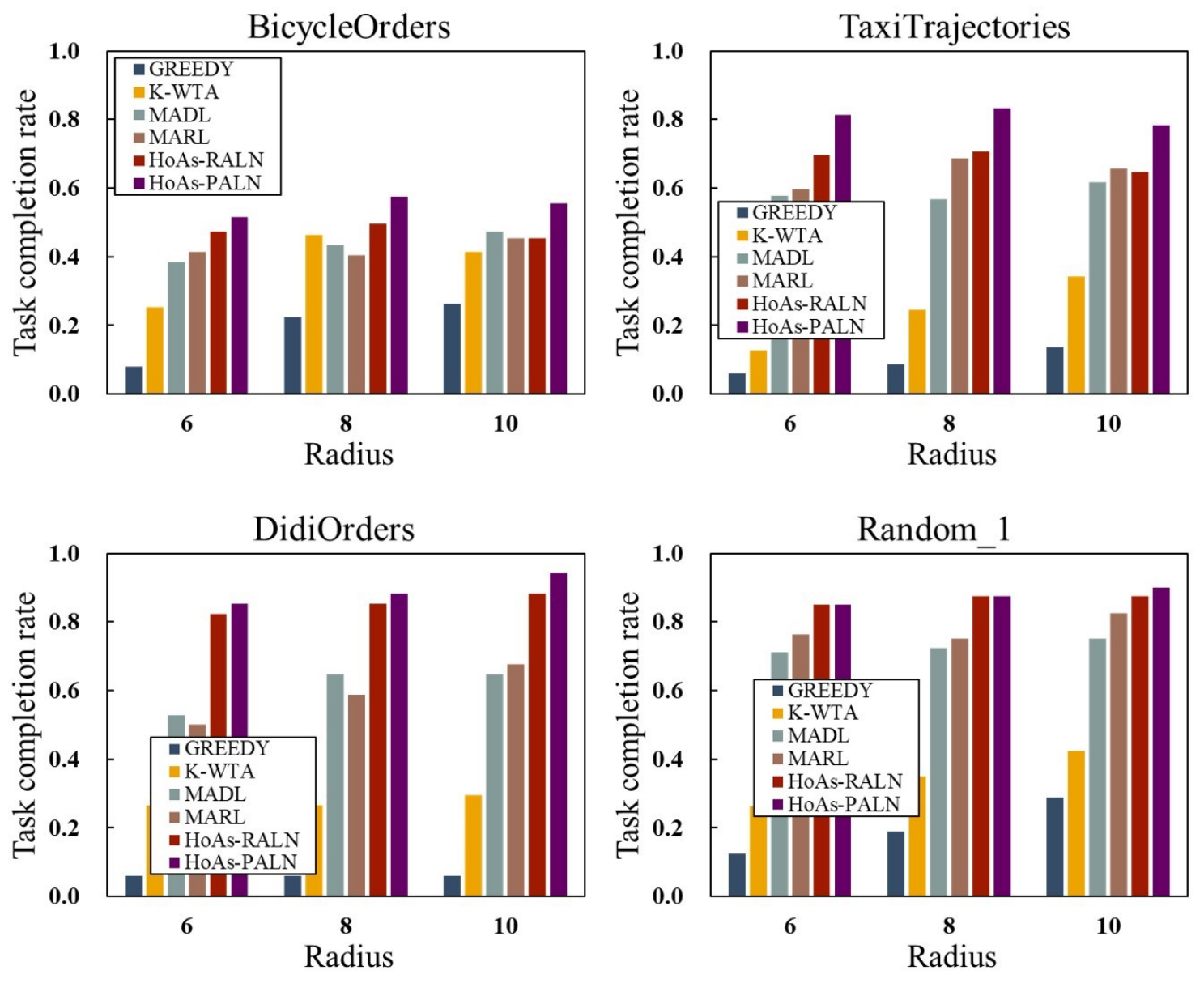}
	\caption{Experimental results with different $Radius$.}
	\label{figure12}
\end{figure}

As $Radius$ increases from 6 to 10, HoAs-PALN consistently outperforms all baselines, see Fig. \ref{figure12} . Specifically, at ranges of 6, 8, and 10, HoAs-PALN achieves task completion rates of 75.80 \%, 79.16 \%, and 79.53 \%, representing improvements of 67.71 \%, 65.24 \%, and 60.88 \% over GREEDY; 53.12 \%, 46.05 \%, and 42.62 \% over K-WTA; 20.70 \%, 19.79 \%, and 17.30 \% over MADL; and 18.93 \%, 18.45 \%, and 14.21 \% over MARL, respectively. These results show that, although task completion rates rise with larger communication ranges, HoAs-PALN maintains a clear advantage under all conditions.

\subsection{Analysis and Discussion}
\label{section 5.5}

\subsubsection*{(1) Simulation trajectory analysis of multiple agents}

We constructed a 3D urban ruins scenario using the Unity engine \cite{unity}, in which the initial positions of 10 unmanned aerial vehicles (UAVs), 15 workers, and 10 vehicles were configured. The autonomous collaborative scheduling of these heterogeneous agents was carried out based on the proposed algorithm HoAs-PALN. As shown in Fig. \ref{figure13}, simulation results show that the trajectories of the UAVs, vehicles, and workers exhibit significant spatial and temporal overlap, indicating a high level of interaction and coordination. These findings suggest that HoAs-PALN effectively facilitates collaborative behavior among the agents, thereby improving task efficiency and overall system performance.

\begin{figure}[ht]
	\setlength{\abovecaptionskip}{0.2cm}
	\setlength{\belowcaptionskip}{-0.25cm}
	\centering 
	\includegraphics[width=\linewidth]{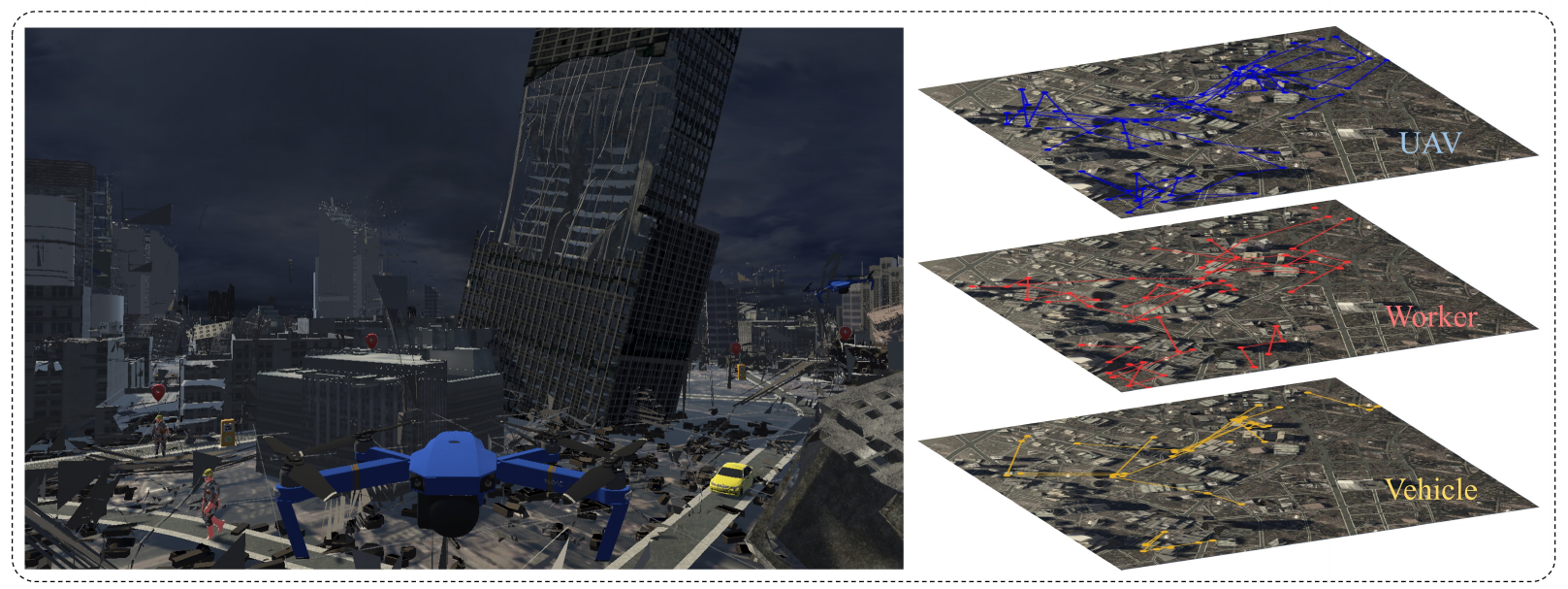}
	\caption{Multi-agent coordination in 3D urban disaster scenarios.}
	\label{figure13}
\end{figure}

\subsubsection*{(2) Influence of coupling strength on task completion rate}

\begin{table*}[ht]
	\centering
	\footnotesize
	\caption{Performance under different coupling strengths.}
	\label{table3}
	\renewcommand{\arraystretch}{1.5}
	\setlength{\tabcolsep}{2mm}
	\begin{tabularx}{\textwidth}{c|c|*{5}{Y}}
		\hline
		\multirow{4}{*}{BicycleOrder} 
		& Coupling strength & 0.5629 & 0.6731 & 0.7244 & 0.8766 & 0.9119 \\ \cline{2-7}
		& Task completion rate & 56.57\% & 55.56\% & 55.56\% & 54.55\% & 49.49\% \\ \cline{2-7}
		& Average time of single decision & 3.11s & 8.55s & 11.22s & 15.99s & 20.03s \\ \cline{2-7}
		& Global nash equilibrium task completion rate & 56.57\% & 55.56\% & 56.57\% & 55.56\% & 55.56\% \\ \cline{2-7}
		\hline
		
		\multirow{4}{*}{TaxiTrajectories} 
		& Coupling strength & 0.2414 & 0.2930 & 0.3382 & 0.3500 & 0.3951 \\ \cline{2-7}
		& Task completion rate & 83.33\% & 81.37\% & 77.45\% & 72.55\% & 62.75\% \\ \cline{2-7}
		& Average time of single decision & 1.72s & 2.21s & 3.33s & 3.99s & 4.05s \\ \cline{2-7}
		& Global nash equilibrium task completion rate & 84.31\% & 84.31\% & 80.39\% & 77.45\% & 67.65\% \\ \cline{2-7}
		\hline
		
		\multirow{4}{*}{DidiOrders} 
		& Coupling strength & 0.4594 & 0.4792 & 0.5433 & 0.5551 & 0.6165 \\ \cline{2-7}
		& Task completion rate & 91.18\% & 88.24\% & 82.35\% & 82.35\% & 70.59\% \\ \cline{2-7}
		& Average time of single decision & 0.31s & 0.44s & 0.46s & 0.52s & 0.50s \\ \cline{2-7}
		& Global nash equilibrium task completion rate & 91.18\% & 91.18\% & 85.29\% & 88.24\% & 79.41\% \\ \cline{2-7}
		\hline
		
		\multirow{4}{*}{Random\_1} 
		& Coupling strength & 0.1342 & 0.1921 & 0.2979 & 0.3721 & 0.4735 \\ \cline{2-7}
		& Task completion rate & 93.75\% & 93.75\% & 60.00\% & 57.50\% & 48.75\% \\ \cline{2-7}
		& Average time of single decision & 1.13s & 1.76s & 1.47s & 1.66s & 1.56s \\ \cline{2-7}
		& Global nash equilibrium task completion rate & 93.75\% & 95.00\% & 63.75\% & 60.00\% & 50.00\% \\ \cline{2-7}
		\hline
	\end{tabularx}
\end{table*}

Based on the analysis of the results in Table \ref{table3}, it can be observed that as the coupling strength increases, the overlap area of the agent communication range expands. As a result, determining the local Nash equilibrium requires coordinating the actions of more neighboring agents, which increases the complexity of the game-theoretic interactions. Consequently, the time required for a single self-scheduling decision to converge to the local Nash equilibrium solution also increases. At the same time, the completion rate of the sensing task is also influenced by the change in coupling strength. It can be seen that as the coupling strength decreases, the sensing task completion rate obtained by HoAs-PALN becomes closer to the global Nash equilibrium solution, which corresponds to the \textbf{Lemma 3} we proposed.

\subsubsection*{(3) Analysis of random behavior choice and probabilistic behavior choice}
\begin{table*}[ht]
	\centering
	\footnotesize
	\caption{Performance comparison between HoAs-RALN and HoAs-PALN.}
	\label{table4}
	\renewcommand{\arraystretch}{1.5}
	\setlength{\tabcolsep}{2mm}
	\begin{tabularx}{\textwidth}{c|*{3}{Y}|*{3}{Y}}
		\hline
		\multirow{3}{*}{\makecell[c]{Datasets}}
		& \multicolumn{3}{c|}{HoAs-RALN} 
		& \multicolumn{3}{c}{HoAs-PALN} \\ \cline{2-7}
		& \makecell{Task\\completion\\rate}
		& \makecell{Average\\time\\of single\\decision}
		& \makecell{Average\\moving distance\\of agent}
		& \makecell{Task\\completion\\rate}
		& \makecell{Average\\time\\of single\\decision}
		& \makecell{Average\\moving distance\\of agent} \\ \hline
		BicycleOrder     & 49.49\% & 25.13s & 3.8965 & 57.58\% & 8.31s & 3.2519 \\ \cline{1-7}
		TaxiTrajectories & 70.59\% & 10.48s & 4.9849 & 83.33\% & 5.16s & 4.4124 \\ \cline{1-7}
		DidiOrders       & 85.29\% & 0.75s  & 3.3313 & 88.24\% & 0.44s & 3.0280 \\ \cline{1-7}
		Random\_1        & 87.50\% & 4.32s  & 7.0031 & 87.50\% & 1.13s & 6.7792 \\ \hline
	\end{tabularx}
\end{table*}
Based on the results in Table \ref{table4}, it can be observed that HoAs-PALN, guided by the probabilistic behavior selection mechanism using the Softmax function, significantly outperforms the pure random strategy of HoAs-RALN in terms of task completion rate, agent decision efficiency, and agent movement distance. This is because the Softmax function amplifies the negative correlation between ``distance'' and ``reward'' exponentially, thereby prioritizing the exploration of nearby high-reward behavior points. This accelerates convergence to local Nash equilibrium solutions. Additionally, the agent tends to select behavior points that are closer, reducing unnecessary movement distance and thereby allowing more time to be spent on task completion, ultimately leading to an improvement in the overall task completion rate.  

\subsubsection*{(4) Effect analysis of dimension reduction in matching process}

\begin{table*}[ht]
	\centering
	\footnotesize
	\caption{Effect analysis of dimension reduction in matching process.}
	\label{table5}
	\renewcommand{\arraystretch}{1.5}
	\setlength{\tabcolsep}{2mm}
	\begin{tabularx}{\textwidth}{p{0.3\textwidth}|*{4}{Y}}
		\hline
		{} 
		& BicycleOrder 
		& TaxiTrajectories 
		& DidiOrders 
		& Random\_1 \\ \cline{1-5}
		\makecell[{{p{0.3\textwidth}}}]{Average time of single decision of agent \textbf{before} dimension reduction} 
		& 331.11s 
		& 215.78s  
		& 0.55s 
		& 3.97s \\ \cline{1-5}
		\makecell[{{p{0.3\textwidth}}}]{Average time of single decision of agent \textbf{after} dimension reduction}  
		& 8.31s   
		& 5.16s    
		& 0.44s 
		& 1.13s \\ \cline{1-5}
		\hline
	\end{tabularx}
\end{table*}

Based on the results presented in Table \ref{table5}, it can be observed that reducing the five-dimensional matching (UAV-worker-vehicle-task point-charge point) to a three-dimensional matching (“UAV-worker-task point” or “UAV-vehicle-charge point”) effectively reduces the complexity and computational load of the matching process, thereby significantly lowering the average decision-making time for each agent. Further analysis of the results derived from the DidiOrders, Random\_1, TaxiTrajectories, and BicycleOrder datasets reveals that as the data scale increases and tasks become more densely distributed, the dimensionality reduction in the matching process increasingly contributes to reducing the average decision-making time of the agents. Thus, it is evident that this dimensionality reduction method holds great potential for practical applications, particularly in large-scale scenarios such as post-disaster environments, where a large number of agents and taskts are involved, making it highly scalable.

\section{Conclusion and Future Work}
\label{section 6}

\subsection{Conclusion}
\label{section 6.1}
Frequent natural disasters pose a significant threat to human life and property. Timely and efficient acquisition of post-disaster environmental information is crucial for the effective implementation of disaster relief operations. In this paper, to efficiently collect post-disaster environmental information, we propose a multi-agent online collaborative scheduling algorithm, HoAs-PALN, based on a matching process adaptive dimensionality reduction and local Nash equilibrium game. The design of HoAs-PALN is as follows: (1) The five-dimensional matching process of “UAV-worker-vehicle-task point-charge point” is reduced to two types of three-dimensional matching processes based on the expected benefits of the UAVs, significantly reducing the scheduling decision time for each agent; (2) The Softmax function is used to convert distance into behavior selection probability, giving higher selection probabilities to superior actions to enhance the exploration efficiency of agents and improve the convergence speed of the algorithm; (3) A local Nash equilibrium state determination method for agents is proposed to ensure that the system scheduling results allow each agent to reach a Nash equilibrium state within its communication coverage. Finally, we conducted extensive experiments using both real and simulated data. The results show that the task completion rate of HoAs-PALN significantly outperforms other baseline methods. Compared with the GREEDY, K-WTA, MADL, and MARL algorithms, the task completion rate of HoAs-PALN is improved by 64.12\%, 46.48\%, 16.55\%, and 14.03\%, respectively, and the decision time for each iteration does not exceed 10 seconds. The findings of this paper provide a feasible technical solution for real-time environmental sensing in post-disaster relief operations. Through the collaborative scheduling of heterogeneous agents, key information from disaster-stricken areas can be quickly obtained, supporting the scientific allocation of rescue resources and the optimization of rescue route planning, which is of significant importance for enhancing post-disaster response efficiency.

\subsection{Future Work}
\label{section 6.2}
Building upon the findings presented in this paper, several aspects merit further investigation in future work: (1) The current study is based on the information exchange capability of each agent’s communication range. Future work should consider introducing communication UAVs to provide a mobile communication network for the agents within a local area network, enabling collaborative scheduling among communication UAVs, sensing UAVs, workers, and vehicles; (2) The current study assumes that UAVs possess ``sensor-computation integration'' capabilities for data collection and preliminary computation. However, their onboard computing resources are still limited by endurance and hardware performance. Future research should investigate dynamic data offloading mechanisms (e.g. edge server collaboration) to offload high-load computation tasks to ground servers or the cloud, balancing computing efficiency and energy consumption, and realizing the collaborative scheduling of sensing, transmission, and computation.

\section*{Acknowledgment}

This work was supported by the China Postdoctoral Science Foundation under Grant Number 2024M762554, the Postdoctoral Fellowship Program (Grade C) of China Postdoctoral Science Foundation under Grant Number GZC20241315, and the Fundamental Research Funds for the Central Universities under Grant Number ZYTS25078.

\bibliographystyle{IEEEtran}
\bibliography{refer_gyt}{}

\begin{IEEEbiography}[{\includegraphics[width=1in,height=1.25in,clip,keepaspectratio]{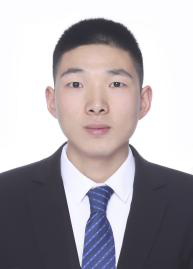}}]{Lei Han}
	received the Ph.D. degree in computer science from Northwestern Polytechnical University, Xi’an, China, in 2023. He currently holds the position of Postdoctoral Researcher at Xidian University. His research interests include ubiquitous computing, mobile crowd sensing, and data mining.
\end{IEEEbiography}

\begin{IEEEbiography}[{\includegraphics[width=1in,height=1.25in,clip,keepaspectratio]{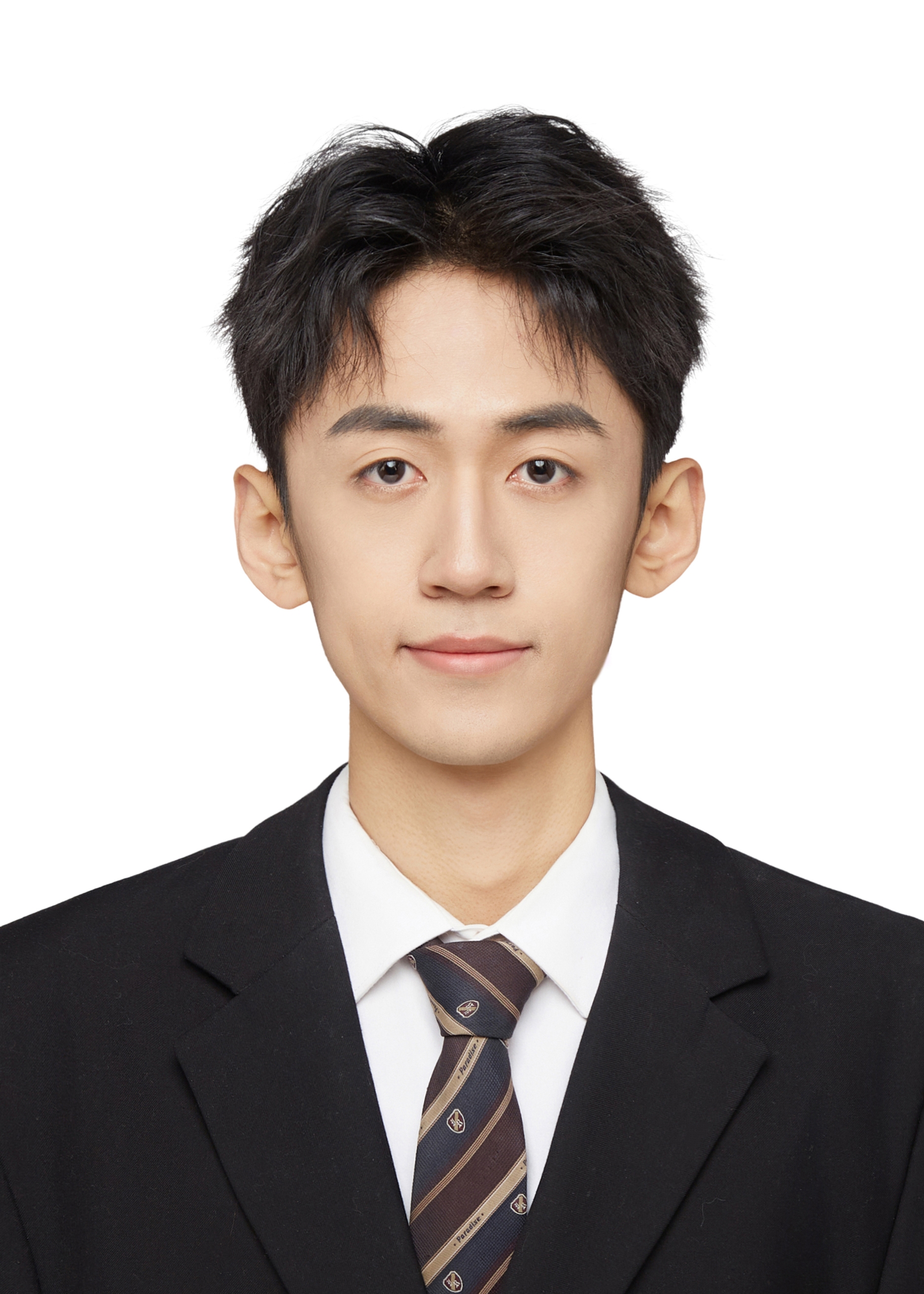}}]{Yitong Guo}
	received the B.Sc. degree in Internet of Things Engineering from Xidian University, Xi'an, China in 2023. He is currently pursuing an M.Sc. degree in electronic information at Xidian University. His research interests include multi-agent system, mobile crowd sensing, and ubiquitous computing.
\end{IEEEbiography}

\begin{IEEEbiography}[{\includegraphics[width=1in,height=1.25in,clip,keepaspectratio]{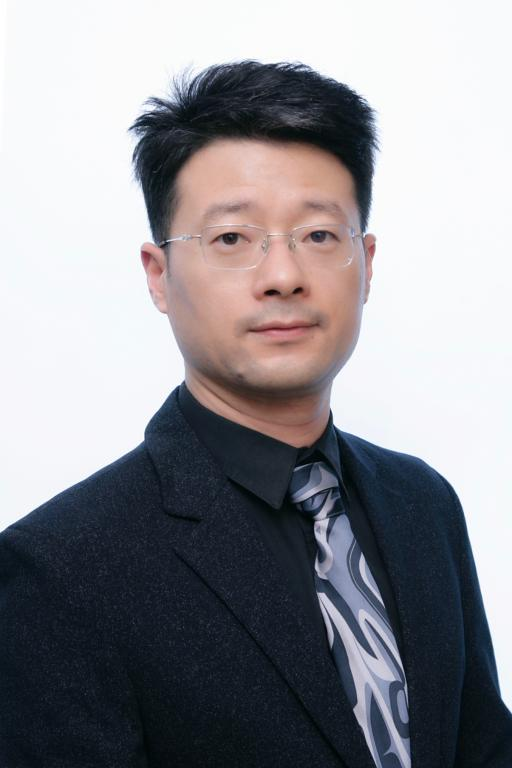}}]{Pengfei Yang}
	received the B.Sc., M.Sc., and Ph.D. degrees from Xidian University, Xi’an, China. He has been an academic visitor for one year in the University of Leeds, UK. He is currently a lecturer in Xidian University. His current research interests include embedded system architecture, memory security, and heterogeneous parallel computing. He is a member of the IEEE.
\end{IEEEbiography}


\begin{IEEEbiography}[{\includegraphics[width=1in,height=1.25in,clip,keepaspectratio]{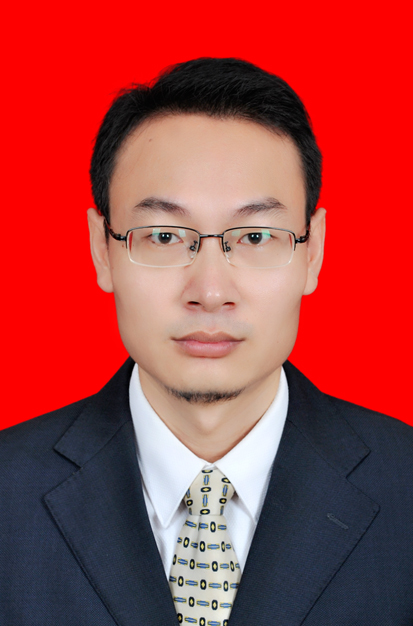}}]{Zhiyong Yu}
	received the M.E. and Ph.D. degrees in computer science and technology from Northwestern Polytechnical University, Xi’an, China in 2007 and 2011, respectively. He is an associate professor in the College of Mathematics and Computer Science, Fuzhou University, Fuzhou, China. He was also a visiting student at Kyoto University, Kyoto, Japan, from 2007 to 2009 and a visiting researcher at the Institut Mines-Telecom, TELECOM SudParis, Evry, France, from 2012 to 2013. His current research interests include pervasive computing, mobile social networks, and mobile crowd sensing.
\end{IEEEbiography}


\begin{IEEEbiography}[{\includegraphics[width=1in,height=1.25in,clip,keepaspectratio]{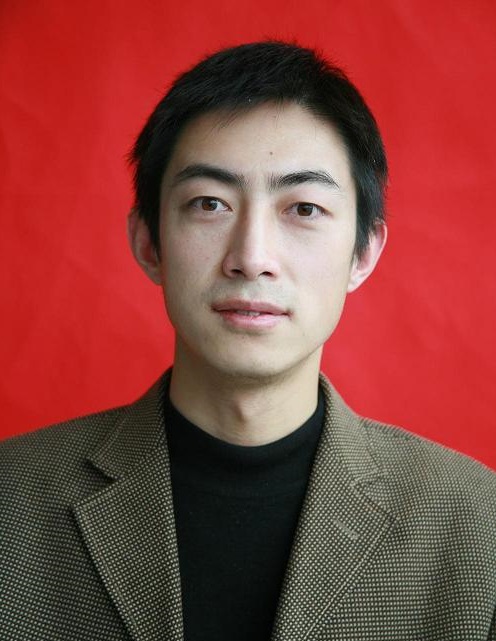}}]{Liang Wang}
	received the Ph.D. degree in computer science from Shenyang Institute of Automation (SIA), Chinese Academy of Sciences, Shenyang, China, in 2014. He is currently an Associate Professor with Northwestern Polytechnical	University, Xi’an, China. His research interests include ubiquitous computing, mobile crowd sensing, and data mining.
\end{IEEEbiography}


\begin{IEEEbiography}[{\includegraphics[width=1in,height=1.25in,clip,keepaspectratio]{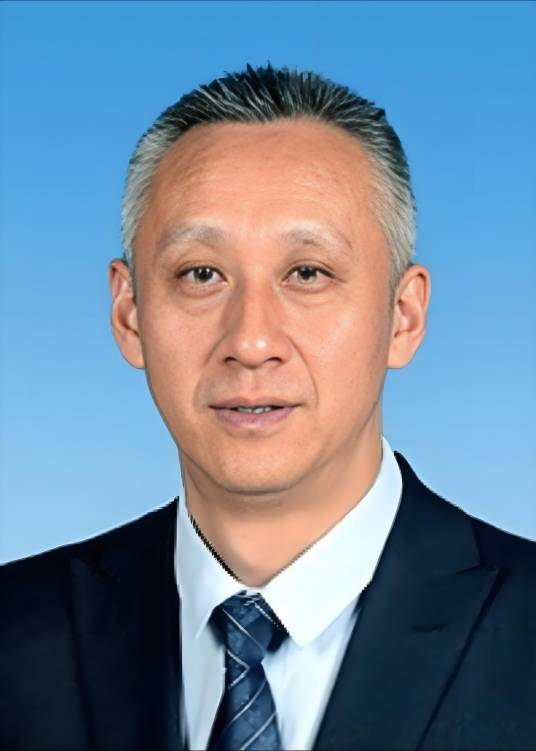}}]{Quan Wang}
	received the B.Sc., M.Sc., and Ph.D. degrees from Xidian University, Xi'an, China. He is currently a professor in Xidian University. His current research interests include input and output technologies and systems, heterogeneous parallel computing, image processing, and image understanding.
\end{IEEEbiography}


\begin{IEEEbiography}[{\includegraphics[width=1in,height=1.25in,clip,keepaspectratio]{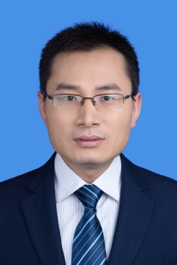}}]{Zhiwen Yu}
	received the M.E. and Ph.D. degrees from Northwestern Polytechnical University, Xi’an, China in 2003 and 2005, respectively. He is a professor in Harbin Engineering University. He visited the Institute of Information and Communication in Singapore from 2004 to 2005. From 2006 to 2009, he was a postdoctoral researcher at Nagoya University and a special researcher at Kyoto University  in Japan. From November 2009 to October 2010, he was funded by the German Humboldt Foundation and went to the University of Mannheim in Germany for collaborative research. His current research interests include pervasive computing, mobile crowdsensing, internet of things, and intelligent information technology.
\end{IEEEbiography}

\end{document}